\documentclass[11pt,letter]{article}

\usepackage[latin1,utf8]{inputenc} 

\usepackage[french,english]{babel}

\usepackage{fullpage}
\usepackage[top=1in, bottom=1in, left=1in, right=1in]{geometry}
\usepackage{amsmath}
\usepackage{amssymb}
\usepackage{amsthm}
\usepackage{mathabx}
\usepackage{graphicx}
\usepackage{wrapfig}
\usepackage{psfrag}
\usepackage{color}

\usepackage[colorlinks=true,linkcolor=blue,citecolor=blue,urlcolor=red]{hyperref}
\usepackage[ruled,vlined,linesnumbered]{algorithm2e}

\usepackage[scriptsize]{subfigure}

\usepackage{tikz}
\usepackage{tkz-graph}
\usetikzlibrary{plotmarks,trees,arrows}
\usetikzlibrary{calc}
\usetikzlibrary{graphs}
\usetikzlibrary{shapes.geometric}
\usepackage{pgf}

\newtheorem{definition}{Definition}

\newtheorem{theorem}{Theorem}

\newtheorem{lemma}{Lemma}

\usepackage[footnotesize,figurename=Fig.]{caption}

\usepackage{mathtools}

\newenvironment{smallitemize} {
 \begin{list}{$-$} {\setlength{\parsep}{0pt}
\setlength{\itemsep}{0pt}} } { \end{list} }

\newcommand{\sep}{\hspace{-0.3cm}:\hspace{-0.3cm}}

\newcommand{\AlgoCLE}{\mbox{\bf{C-LE}}}

\newcommand{\ID}{\mbox{\sc id}}
\newcommand{\T}{\mbox{\tt T}}

\newcommand{\p}{\mbox{\sf p}}
\newcommand{\kid}{\mbox{\sf k}}
\newcommand{\degr}{\mbox{$\delta$}} 

\newcommand{\Child}{\mbox{\tt Ch}}
\newcommand{\Bit}{\mbox{\rm Bit}}

\newcommand{\Phase}{\mbox{\sf ph}}
\newcommand{\Bp}{\mbox{\sf Bp}}
\newcommand{\SB}{\mbox{$\widehat{\sf B}$}}
\newcommand{\PactID}{\mbox{${\mathcal Cid}$}}

\newcommand{\CB}{\mbox{\tt C$\widehat{\sf B}$}}
\newcommand{\CPh}{\mbox{\tt CPh}}
\newcommand{\CBp}{\mbox{\tt CBp}}
\newcommand{\MinCID}{\mbox{\tt MinCid}}
\newcommand{\MaxCID}{\mbox{\tt MaxCid}}

\newcommand{\ErT}{\mbox{\tt ErT}}
\newcommand{\SErB}{\mbox{\tt SErB}}
\newcommand{\TErB}{\mbox{\tt TErB}}

\newcommand{\ResetCID}{\mbox{$\mathcal ResetCid$}}
\newcommand{\SIncPh}{\mbox{$\mathcal IncPh$}}
\newcommand{\TIncPh}{\mbox{$\mathcal IncPh$}}
\newcommand{\Opt}{\mbox{$\mathcal Opt$}}
\newcommand{\AlgoC}{\mbox{\bf{C-Color}}}

\newcommand{\couleur}{\mbox{\sf c}}
\newcommand{\conflit}{\mbox{\sf $\widecheck{c}$}}
\newcommand{\Mc}{\mbox{\sf $\widehat{c}$}}

\newcommand{\MinCf}{\mbox{\tt mC}}
\newcommand{\MaxC}{\mbox{\tt MC}}
\newcommand{\Illicit}{\mbox{\tt Bad}}
\newcommand{\Player}{\mbox{\tt Player}}
\newcommand{\PlayerR}{\mbox{\tt PlayR}}

\newcommand{\RelayC}{\mbox{\tt Relay}}
\newcommand{\RelayUp}{\mbox{\tt RUp}}
\newcommand{\Loser}{\mbox{\tt Loser}}

\newcommand{\Oth}{\mbox{\tt Oth}}

\newcommand{\SIncPhase}{\mbox{\tt SPh$^{+}$}}
\newcommand{\TIncPhase}{\mbox{\tt TPh$^{+}$}}

\newcommand{\Newcolor}{\mbox{$\mathcal Newcolor$}}
\newcommand{\AUpdate}{\mbox{$\mathcal Update$}}

\newcommand{\RDelta}{\mathbb{R}_{\Delta}}
\newcommand{\RDeltaP}{\mathbb{R}_{\Delta}^+}
\newcommand{\RUp}{\mathbb{R}_{\tt Up}}
\newcommand{\RColor}{\mathbb{R}_{\tt Color}}
\newcommand{\RBit}{\mathbb{R}_{\tt Bit}}
\newcommand{\RRelay}{\mathbb{R}_{\tt Relay}}

\newcommand{\AlgoBreakR}{\mbox{\bf{Break}}}

\newcommand{\CAlgoBreakG}{\mbox{\bf{C-Break}}}
\newcommand{\frozen}{\mbox{\sf froz}}
\newcommand{\minId}{\mbox{\sf m}}
\newcommand{\del}{\mbox{\sf E}}
\newcommand{\Active}{\mbox{\tt Active}}

\newcommand{\ErCycle}{\mbox{\rm ErCycle}}
\newcommand{\Improve}{\mbox{\rm Improve}}
\newcommand{\RStart}{\mathbb{R}_{\tt Start}}
\newcommand{\RMin}{\mathbb{R}_{\tt Min}}
\newcommand{\RID}{\mathbb{R}_{\tt ID}}
\newcommand{\RInc}{\mathbb{R}_{\tt Inc}}

\newcommand{\AlgoST}{\mbox{\bf{ST}}}
\newcommand{\AlgoCST}{\mbox{\bf{C-ST}}}

\newcommand{\N}{\mbox{$\overline{\sf n}$}}

\newcommand{\R}{\mbox{\sf R}}
\newcommand{\new}{\mbox{\sf new}}
\newcommand{\chek}{\mbox{\sf check}}
\newcommand{\Cand}{\mbox{$f$}}
\newcommand{\ErST}{\mbox{\rm ErST}}
\newcommand{\RReRoot}{\mathbb{R}_{\tt ReRoot}}
\newcommand{\RDel}{\mathbb{R}_{\tt Del}}
\newcommand{\RMerge}{\mathbb{R}_{\tt Merge}}
\newcommand{\RUpdate}{\mathbb{R}_{\tt Update}}
\newcommand{\RPath}{\mathbb{R}_{\tt Path}}

\newcommand{\AlgoFreeze}{\mbox{\bf{Freeze}}}
\newcommand{\RErCycle}{\mathbb{R}_{\tt Error}}
\newcommand{\RFroze}{\mathbb{R}_{\tt Froze}}
\newcommand{\RPrun}{\mathbb{R}_{\tt Prun}}

\begin{document}

\title{Compact  Self-Stabilizing Leader Election \\ for Arbitrary Networks}
\date{}

\author{
L\'elia Blin\footnote{Sorbonne Universit\'es, UPMC Univ Paris 06, CNRS, Universit\'e d'Evry-Val-d'Essonne, LIP6 UMR 7606, 4 place Jussieu 75005, Paris.}
\and
S\'ebastien Tixeuil\footnote{Sorbonne Universit\'es, UPMC Univ Paris 06, CNRS, LIP6 UMR 7606, 4 place Jussieu 75005, Paris.}
}
\maketitle
\begin{abstract}
  We present a self-stabilizing leader election algorithm for arbitrary networks, with space-complexity $O(\max\{\log \Delta, \log \log n\})$ bits per node in $n$-node networks with maximum degree~$\Delta$. This space complexity is sub-logarithmic in $n$ as long as $\Delta = n^{o(1)}$.  The best space-complexity known so far for arbitrary  networks was $O(\log n)$ bits per node, and algorithms with sub-logarithmic space-complexities were known  for the ring only. To our knowledge, our algorithm is the first algorithm for self-stabilizing leader election to break the $\Omega(\log n)$ bound for silent algorithms in arbitrary networks. Breaking this bound was obtained via the  design of a (non-silent) self-stabilizing algorithm using sophisticated tools such as solving the distance-2 coloring problem in a silent self-stabilizing manner, with space-complexity $O(\max\{\log \Delta, \log \log n\})$ bits per node. Solving this latter coloring problem allows us to implement a sub-logarithmic encoding of spanning trees --- storing the IDs of the neighbors requires $\Omega(\log n)$ bits per node, while we encode spanning trees using $O(\max\{\log \Delta, \log \log n\})$ bits per node. Moreover, we show how to construct such compactly encoded spanning trees without relying on variables encoding distances or number of nodes, as these two types of variables would also require $\Omega(\log n)$ bits per node.
\end{abstract}

\thispagestyle{empty}
\setcounter{page}{0}
\newpage
\section{Introduction}
\label{sec:intro}
\vspace*{-0,3cm}
This paper tackles the problem of designing memory efficient
self-stabilizing algorithms for the leader election
problem. \emph{Self-stabilization}~\cite{D74j,D00b,T09bc} is a general
paradigm to provide recovery capabilities to  networks. 
Intuitively, a protocol is self-stabilizing if it is able to
recover from any transient failure, without external
intervention. \emph{Leader election} is one of the fundamental
building blocks of distributed computing, as it enables a single node
in the network to be distinguished, and thus to perform specific
actions. Leader election is especially important in the context of
self-stabilization as many protocols for various problems assume that
a single leader exists in the network, even after faults occur. Hence, a
self-stabilizing leader election mechanism enables such protocols to
be run in networks where no leader is given a priori, by using simple
stabilization-preserving composition techniques~\cite{D00b}. \emph{Memory
efficiency} relates to the amount of information to be sent to
neighboring nodes for enabling stabilization. A small space-complexity 
induces a smaller amount of information transmission, which (1)~reduces the overhead of self-stabilization when there are no faults, or
after stabilization~\cite{ANT12c}, and (2)~facilitates mixing
self-stabilization and replication~\cite{GCH06c,HP00j}. 

A foundational result regarding  space-complexity in the context of
self-stabilizing silent algorithms\footnote{An algorithm is \emph{silent} if each of
its executions reaches a point in time after which the states of nodes do not change. A non-silent algorithm is said to be \emph{talkative} (see~\cite{BlinT17}).} is due to Dolev et al.~\cite{DGS99j}, stating that, in $n$-node networks, $\Omega(\log n)$ bits of memory
per node are required for solving tasks such as leader
election. So, only \emph{talkative} algorithms
may have $o(\log n)$-bit space-complexity for self-stabilizing leader election.
Several attempts to design compact
self-stabilizing leader election algorithms (i.e., algorithms with space-complexity $o(\log n)$ bits) were performed, but restricted to rings. The algorithms by
Mayer et al.~\cite{MOOY92c}, by Itkis and Levin~\cite{IL94c},
and by Awerbuch and Ostrovsky~\cite{AO94c} use a constant number of
bits per node, but they only guarantee \emph{probabilistic}
self-stabilization (in the Las Vegas sense). 
\emph{Deterministic} self-stabilizing leader election algorithms for
rings were first proposed by Itkis et al.~\cite{ILS95c} for
rings with a prime number of nodes. Beauquier et al.~\cite{BGJ99c} consider 
rings of arbitrary size, but assume that node identifiers in $n$-node rings are
bounded from above by $n+k$, where $k$ is a small constant. A recent result by Blin et
  al.~\cite{BlinT17} demonstrates that both previous constraints in the
deterministic setting are unnecessary, by presenting a deterministic 
self-stabilizing leader election algorithm for rings of arbitrary size 
using identifiers of arbitrary polynomially bounded values, with space complexity  $O(\log \log n)$ bits. 

In general networks, self-stabilizing leader election is tightly
connected to self-stabilizing tree-construction. On the one hand, the
existence of a leader permits time- and memory-efficient
self-stabilizing tree-construction~\cite{ChenYH91,DIM93j,CollinD94,BlinPR13,KormanKM11}. 
On the other hand, growing and merging trees is the main technique for
designing self-stabilizing leader election algorithms in networks, as the leader is often the root of an inward tree~\cite{AfekKY90,AG94j,AB98j,BlinDPR10}.
To the best of our
knowledge, all algorithms that do not assume a pre-existing leader~\cite{AfekKY90,AG94j,AB98j,BlinBD15} for tree-construction use
$\Omega(\log n)$ bits per node. This high space-complexity is due
to the implementation of two main techniques, used by all algorithms, and  recalled below.
 
The first technique is the use of a \emph{pointer-to-neighbor}
variable, that is meant to designate unambiguously one particular
neighbor of every node. For the purpose of tree-construction, pointer-to-neighbor variables are typically used to store the parent node in the
constructed tree.
Specifically, the parent of every node is designated unambiguously by its identifier, requiring
$\Omega(\log n)$ bits for each pointer variable. In principle, it would be
possible to reduce the memory to $O(\log \Delta)$ bits per
pointer variable in networks with maximum degree~$\Delta$, by using node-coloring at distance~2 instead of
 identifiers to identify neighbors. However, this in turn would require the
availability of a self-stabilizing distance-2 node-coloring
algorithm that uses $o(\log n)$ bits per node. Previous
self-stabilizing distance-2 coloring algorithms use variables of large size.
 For instance, in the algorithm by
Herman et al. \cite{HT04c}, every node communicates its distance-3 neighborhood to all its neighbors, which yields a space-complexity of $O(\Delta^3\log n)$ 
bits. Johnen et al.~\cite{GradinariuJ01} draw random colors in
the range $[0,n^2]$, which yields a space-complexity of $O(\log n)$ bits. 
Finally, while  the deterministic algorithm of Blair et
  al.~\cite{BlairM12} reduces the space-complexity to $O(\log \Delta)$ bits per
node, this is achieved by ignoring the cost for storing another pointer-to-neighbor variable at each node. 
In absence of  a distance-2 coloring (which their algorithm~\cite{BlairM12} is precisely supposed to produce), their implementation still requires $\Omega(\log n)$ bits per node. 
To date, no
self-stabilizing algorithm implement pointer-to-neighbor variables with space-complexity $o(\log n)$ bits  in arbitrary networks.

The second technique for tree-construction or leader election is the use of a \emph{distance} variable that is meant to
store the distance of every node to the elected node in the network. Such
distance variable is used in self-stabilizing spanning tree-construction for  breaking cycles resulting from arbitrary initial
 state (see~\cite{AfekKY90,AG94j,AB98j}). Clearly, storing distances in $n$-node networks may require 
$\Omega(\log n)$ bits per node. 
There are a few self-stabilizing tree-construction algorithms that are not using explicit distance variables (see, e.g.,~\cite{BGJD02j,DT03j,DDT06j}), but
their space-complexity is $O(n \log n)$  bits~\cite{DT03j,DDT06j} or $O(\log n + \Delta)$~\cite{BGJD02j}.
Using the general principle of distance variables with space-complexity below  $\Theta(\log n)$ bits was attempted by Awerbuch et
  al.~\cite{AO94c}, and Blin et al.~\cite{BlinT_DISC13}. These papers distribute pieces of information about the distances to 
  the leader among the nodes according to different mechanisms, enabling to store  $o(\log n)$ bits
per node. However, these sophisticated mechanisms  have only
been demonstrated in rings. To date, no
self-stabilizing algorithms implement distance variables with space-complexity $o(\log n)$ bits in arbitrary networks.

\vspace*{-0,3cm}
\subsection*{Our results}
\vspace*{-0,2cm}
In this paper, we design and analyze a self-stabilizing leader election algorithm with space-complexity $O(\max\{\log \Delta, \log \log n\})$ bits in $n$-node networks with maximum degree~$\Delta$. This algorithm is the first self-stabilizing leader election algorithm for arbitrary networks with space-complexity $o(\log n)$ (whenever $\Delta = n^{o(1)}$). It is designed for the standard state model (a.k.a.~sha\-red memory model) for self-stabilizing algorithms in networks, and it performs against the unfair distributed scheduler. 

The design of our algorithm requires to overcome  several  bottlenecks, including the difficulties of manipulating pointer-to-neighbor and distance variables using $o(\log n)$ bits in arbitrary networks. Overcoming these bottlenecks was achieved thanks to the development of sub-routine algorithms, each deserving independent special interest, described hereafter. 

First, we generalize to arbitrary networks the techniques proposed~\cite{BlinT_DISC13,BlinT17} for rings, and aiming at publishing the identifiers in a bit-wise manner. This generalization allows us to manipulate the identifiers with just $O(\log \log n)$ bits of memory per node. 

Second, we propose a \emph{silent}  self-stabilizing  algorithm for distance-2 coloring with space-complexity $O(\max\{\log \Delta, \log \log n\})$ bits. As opposed to previous distance-2 coloring algorithms, we do not use identifiers for encoding pointer-to-neighbor variables, but we use a compact representation of the identifiers to break symmetries. This allows us to design a compact encoding of spanning trees. 

Third, we design a new technique to detect the presence of cycles in the initial configuration resulting from a transient failure. This technique does not use distances, but is based on the uniqueness  of each  identifier in the network. Notably, this technique can be implemented by a \emph{silent} self-stabilizing algorithm, with space-complexity $O(\max\{\log \Delta, \log \log n\})$ bits per node.

Last but not least, we design a new technique to avoid the creation of cycles during the execution of the leader election algorithm. Again, this technique does not uses distances but maintains a spanning forest, which eventually reduces to a single spanning tree rooted at the leader at the completion of the leader election algorithm. Implementing this technique results in a self-stabilizing algorithm with space complexity $O(\max\{\log \Delta, \log \log n\})$ bits per node.

\vspace{-0,3cm}
\section{Model and definitions}
\label{sec:model}
\vspace{-0,2cm}
\subsection{Protocol syntax and semantics} 

We consider a distributed system consisting of $n$ processes that form a
arbitarry communication graph. The processes are represented by the nodes of this graph, and the edges represent pairs of processes
that can communicate directly with each other. Such processes are said to be \emph{neighbors}. 
Let $G=(V,E)$ be an $n$-node  graph, where $V$ is the set of nodes, and $E$ the set of edges and $\Delta$ the degree of the graph. 
A node $v$ has access to a constant unique identifier $\ID_v$, but can only access its identifier one bit at a time, using the $\Bit(x,\ID_v)$ function, which  returns the position of the $x^\mathrm{th}$ most significant bit equal to $1$ in $\ID_v$. This position can be encoded with $O(\log \log n)$ bits when identifiers are encoded using $O(\log n)$ bits, as we assume they are. 
A node $v$ has access to locally unique port numbers associated to its adjacent edges. We do not assume any consistency between port numbers of a given edge. In short, port numbers are constant throughout the execution but initialized by an adversary.
Each process contains
variables and rules. A variable ranges over a   domain of values. The variable $var_{v}$ denote the variable $var$ located at node $v$. 
A rule is of the form $\langle label \rangle : \langle guard \rangle\longrightarrow \langle command \rangle.$
  A \emph{guard} is a boolean
predicate over process variables. A \emph{command} is a set of
variable-assignments. 
A command of process $p$ can only update its own variables.  
On the other hand, $p$ can read the variables of its neighbors. 
This classical communication model is called the {\em state model} or 
the \emph{state-sharing communication model}.

An assignment of values to all variables in the system is called a
\emph{configuration}. A rule whose guard is \textbf{true} in some
system configuration is said to be \emph{enabled} in this configuration. The rule
is \emph{disabled} otherwise. The atomic execution of a subset of
enabled rules (at most one rule per process) results in a transition of the system from one configuration to
another. This transition is called a \emph{step}. A \emph{run}
of a distributed system is a maximal alternating sequence of configurations and steps. 
Maximality means that the execution is either infinite, or its final configuration 
has no rule enabled. 
\vspace{-0,3cm}
\subsection{Schedulers}

The asynchronism of the system is modeled by an adversary (a.k.a. \emph{scheduler}) that chooses, at each step, the subset of enabled processes that are allowed to execute one of their rules during this step. The literature proposed a lot of daemons depending of their characteristics (like fairness, distribution, ...), see~\cite{DT11r} for a taxonomy of these scheduler. Note that we assume here an unfair distributed scheduler. This scheduler is the most challenging since  no assumption is made of the subset of enabled processes chosen by the scheduler at each step (That only require this set to be non empty if the set of enabled processes is not empty in order to guarantee progress of the algorithm.)

\vspace{-0,3cm}
\subsection{Predicates and specifications}
 
A predicate is a boolean function over  configurations. A
configuration \emph{conforms} to some predicate $R$, if $R$ evaluates
to \textbf{true} in this configuration. The configuration
\emph{violates} the predicate otherwise.  Predicate $R$ is
\emph{closed} in a certain protocol $P$, if every configuration of
a run of $P$ conforms to $R$, provided that the protocol starts
from a configuration conforming to $R$. Note that if a protocol
configuration conforms to $R$, and the  configuration resulting from 
the execution of any step of $P$ also conforms to $R$, then $R$
is closed in $P$.


\emph{Problem specification} prescribes the protocol behavior. The output of the protocol is carried through external variables, that are updated by the protocol, and used to display the results of the protocol computation. The problem specification is the set of sequences of configurations of external variables. 

A protocol implements the specification. Part of the implementation is the mapping from the protocol configurations to the specification configurations. This mapping does not have to be one-to-one. However, we only consider unambiguous protocols where each protocol configuration maps to only one specification configuration. Once the mapping between protocol and specification configurations is established, the protocol runs are mapped to specification sequences as follows. Each protocol configuration is mapped to the corresponding specification configuration. Then, stuttering, the consequent identical specification configurations, is eliminated. 
Overall, a run of the protocol satisfies the specification if its mapping belongs to the specification. Protocol $P$ \emph{solves} problem $S$ under a certain scheduler if every run of $P$ produced by that scheduler satisfies the specifications defined by $S$. 
Given two predicates $l_1$ and $l_2$ for protocol $P$, $l_2$ is an \emph{attractor} for $l_1$ if every run that
starts from a configuration that conforms to $l_1$ contains a
configuration that conforms to $l_2$. Such a relationship is denoted
 by $l_1 \triangleright l_2.$ Also, the $\triangleright$ relation is transitive: if $l_1$, $l_2$, and $l_3$ are predicates for $P$, and $l_1 \triangleright l_2$ and $l_2 \triangleright l_3$, then $l_1 \triangleright l_3$. In this last case, $l_2$ is called an \emph{intermediate} attractor towards $l_{3}$.

\begin{definition}[Self-stabilization]
A protocol $P$ is \emph{self-stabilizing} \cite{D74j} to specification
$S$ if there exists a predicate $L$ for $P$ such that:
\begin{enumerate}
\vspace{-0,2cm}
\setlength\itemsep{-0.3em}
\item $L$ is an attractor for \emph{true},
\item Any run of $P$ starting from a configuration satisfying $L$ satisfies $S$.
\end{enumerate}
\end{definition}
\begin{definition}[Leader Election]
Consider a system of processes where each process' set of variables is mapped to a boolean specification
variable leader denoted by $\ell$. 
The \emph{leader election} specification sequence consists in a single specification configuration where a unique process $p$ maps to $\ell_{p}=true$, and 
every other process $q\neq p$ maps to $\ell_q=false$. 
\end{definition}

\vspace*{-0,3cm}
\section{Compact  self-stabilizing leader election for  networks} 
\vspace*{-0,2cm}
Our self-stabilizing leader election algorithm is based on a spanning tree-construction rooted at a maximum degree node, without using distances. If multiple maximum degree nodes are present in the network, we break ties with colors and if necessary with identifiers. 

\begin{theorem}
\label{theo:CompoC}
Algorithm \AlgoCLE\/ solves the leader election problem  in a talkative self-stabilizing manner in any $n$-node graph, assuming the state model and a distributed unfair scheduler, with $O(\max\{\log \Delta, \log \log n\})$ bits of memory per node, where $\Delta$ is the graph's degree.
\end{theorem}

Our talkative self-stabilizing algorithm reuses and extends a technique for obtaining compact identifiers of size $O(\log \log n)$ bits per node presented in Section~\ref{sub:Compact}. Then, the leader election process consists in running several algorithms layers using decreasing priorities: \begin{enumerate}
\setlength\itemsep{-0.1em}
\item A silent self-stabilizing distance-2 coloring presented in subsection~\ref{sub:Color} that permits to implement pointer-to-neighbors with $o(\log n)$ bits per node.
  \item A silent self-stabilizing cycle and illegitimate sub spanning tree destruction reused from previous work~\cite{BlinF15,BlinT17} presented in subsection~\ref{sub:Freeze}.  
\item A silent self-stabilizing cycle detection that does not use distance to the root variables  presented in subsection~\ref{sub:Break}. 
\item  A talkative self-stabilizing spanning tree-construction, that still does not use distance to the root variables, presented in subsection~\ref{sub:LE}. This algorithm is trivially modified to obtain a leader election algorithm.
\end{enumerate}
Due to the lack of space most of the proofs and predicates are delegated to the Appendix.
\vspace*{-0,3cm}
\subsection{Compact memory using identifiers}
\label{sub:Compact}

As many deterministic self-stabilizing leader election algorithms, our approach ends up comparing node unique identifiers. However, to avoid communicating the full $\Omega(\log n)$ bits to each neighbor at any given time, we reuse the scheme devised in previous work~\cite{BlinT_DISC13,BlinT17} to progressively publish node identifiers.
Let $\ID_{v}$ be the identifier of node $v$. We assume that  $\ID_v=\sum_{i=0}^k b_i 2^i$.
Let $I_{v}=\big \{i\in\{0,...,k\},b_i\neq 0\big\}$ be the set of all non-zero bit-positions in the binary representation of $\ID_v$. Then, $I_{v}$ can be written as $\{pos_1,...,pos_j\}$, where $pos_{k}>pos_{k+1}$.
In the process of comparing node unique identifiers during the leader election algorithm execution, the nodes must first agree on the same bit-position $pos_{j-i+1}$ (for $i=1,\dots,j$); this step of the algorithm defines \emph{phase}~$i$. Put differently, the bit-positions are communicated in decreasing order of significance in the encoding of the identifier. 

If all identifiers are in $[1,n^c]$, for some constant $c\geq 1$, then the communicated bit-positions are less than or equal to $c\lceil\log n\rceil$, and thus can be represented with $O(\log\log n)$ bits. However, the number of bits used to encode identifiers may be different for two given nodes, so there is no common upper bound for the size of identifiers. 
%
Instead, we use variable $\SB_{v}$, which represents the most significant bit-position of node $v$. In other words, $\SB_{v}$ represents the size of the binary representation of $\ID_{v}$. The variables $\Phase$, $\Bp$ are the core of the identifier comparison process. Variable $\Phase_v$ stores the current phase number $i$, while Variable $\Bp_v$ stores the bit-position of $\ID_v$ at phase $i$. Remark that the number of non-zero bits can be smaller than the size of the binary representation of the identifier of the node, so if there are no more non-zero bit at phase $i\leq \SB_{v}$, we use $\Bp_{v}=-1$.
To make the algorithm more readable, we introduce Variable $\PactID_{v}=(\SB_{v},\Phase_{v},\Bp_{v})$, called a \emph{compact identifier} in the sequel. When meaningful, we use $\PactID_v^{\,i}=(\SB_{v},\Bp_{v})$, where $i=\Phase_{v}$.

Node $v$ can trivially detect an error (see predicate $\ErT(v)$) whenever its compact identifier does not match its global identifier, or its phase is greater than $\SB_v$.
Moreover, the phases of neighboring nodes must be close enough: a node's phase may not be more than $1$ ahead or behind any of its neighbors; also a node may not have a neighbor ahead and another behind. 
Predicate $\SErB(v,S)$ captures these conditions, where $S(v)$ denotes a subset of neighbors of $v$. The set $S$ should be understood as an input provided by an upper layer algorithm. If $v$ detects an error through $\ErT(v)$ or $\SErB(v,S)$, it resets its compact identifier to its first phase value (see command $\ResetCID(v)$).
The compact identifier of $u$ is smaller (respectively greater) than the compact identifier of $v$, if the most significant bit-position of $u$ is smaller (respectively greater) than the  most significant bit-position of $v$, or if the most significant bit-position of $u$ is equal to the  most significant bit-position of $v$, $u$ and $v$ are in the same phase, and the bit-position of $u$ is smaller (respectively greater) than the bit-position of $v$:
\vspace{-0,2cm}
\begin{equation}
 \PactID_u^{\,i}<_{c}\PactID_v^{\,i}\equiv(\SB_{u}<\SB_{v}) \vee   \big((\SB_{v}=\SB_{u}) \wedge (\Bp_{u}<\Bp_{v})\big) 
\label{P:Inf}
\vspace{-0,2cm}
\end{equation}
When two nodes $u$ and $v$ have the same most significant bit-position and the same bit position at phase $i<\SB_{v}$, they are possibly equal with respect to compact identifiers (denoted by $\simeq_c$).
\vspace{-0,2cm}
\begin{equation}
\begin{split}
\PactID_u^{\,i}\simeq_{c}\PactID_v^{\,i}\equiv    (i<\SB_v) \wedge\big((\SB_{v}=\SB_{u}) \wedge  (\Bp_{u}=\Bp_{v})\big) 
\label{P:Meq}
\end{split}
\vspace{-0,3cm}
\end{equation}
Finally, two nodes $u$ and $v$ have the same compact identifier (denoted by $=_c$) if their phase reaches the size of the binary representation of the identifier of the two nodes, and their last bit-position is the same.
\vspace{-0,3cm}
\begin{equation}
\begin{split}
\PactID_u=_{c}\PactID_v\equiv (i=\SB_{v}=\SB_{u}) \wedge (\Bp_{u}=\Bp_{v})
\label{P:Eq}
\end{split}
\vspace{-0,8cm}
\end{equation}
The predicates $\SIncPhase(v)$ and $\TIncPhase(v,S)$ check if a node $v$ can increases its phase (or restarts $\PactID_v$), the first one is dedicated to the silent protocols, the second one is dedicated to  the talkative protocols. The command $\TIncPh(v)$ is  dedicated for increasing phases or restarting $\PactID_v$.
Last, the command $\Opt$  assigns at a node $v$ the minimum (or maximum) compact identifier in the subset  of neighbors $S(v)$. We have now, all the principals ingredients to use compact identifiers.

\vspace*{-0,3cm}
\subsection{Silent self-stabilizing distance-2 coloring}
\label{sub:Color}

In this section, we provide a solution to assign colors that are unique up to distance two (and bounded by a polynom of the graph degree) in any  graph. Those colors are meant to efficiently implement the pointer-to-neighbor mechanism that otherwise requires $\Omega(\log n)$ bits per node.

Our solution uses compact identifiers to reduce memory usage.
When a node $v$ has the same color as (at least one of) its neighbors, then if the node $v$ has the smallest conflicting color in its neighborhood and is not the biggest identifier among conflicting nodes, then $v$ changes its color. To make sure a fresh color is chosen by $v$, all nodes publish the maximum color used by their neighborhood (including themself). So, when $v$ changes its color, it takes the maximum advertised color plus one. Conflicts at distance two are resolved as follow: let us consider two nodes $u$ and $v$ in conflict at distance two, and let $w$ be (one of) their common neighbor; as $w$ publishes the color of $u$ and $v$, it also plays the role of a relay, that is, $w$ computes and advertises the maximum identifiers between $u$ and $v$, using the compact identifiers mechanisms that were presented above; a bit by bit, then, if $v$ has the smallest identifier, it changes its color to a fresh one. 
To avoid using too many colors when selecting a fresh one, all changes of colors are made modulo an upper bound on the number of neighbors at distance 2, which is computed locally by each node.

\vspace*{-0,3cm}
\subsubsection{Self-stabilizing algorithm description}
\vspace*{-0,1cm}
Each node $v$ maintains a color variable denoted by $\couleur_v$ and a degree variable  denoted by $\degr_v$.
A variable $\conflit_{v}$ stores the minimum color in conflict in its neighborhood (including itself).
The variable $\Mc_{v}$ stores the maximum color observed in its neighborhood.
We call $v$ a \emph{player} node when $v$ has the minimum color in conflict. Also, we call $u$ a \emph{relay} node when $u$ does not have the minimum color in conflict, yet at least two of its neighbors have the minimum color in conflict.

The rule $\RDelta$ assures that the degree variable is equal to the degree of the node.
Each node $v$ must maintain its color in  range $[1,\Delta(v)^{2}+1]$ to satisfy the memory requirements of our protocol, where $\Delta(v)$ is a function that returns the maximum degree of its neighborhood (including itself). Whenever $v$'s color exceeds its expected range,  rule $\RDeltaP$ resets the color to one.
Rule $\RUp$ is dedicated to updating the variables of $v$ whenever they do not match the observed neighborhood of $v$ (see $\Illicit(v)$), or when a player node has an erroneous phase variable when comparing its identifier with another player node (see function $\Oth(v)$). In both cases, the $v$ computes the minimum and maximum color and resets its compact identifier variable (see command $\AUpdate(v)$).
The rule $\RColor$ increases the color of the node $v$ but maintains the color in some range (see command $\Newcolor(v)$), when $v$ has the minimum color in conflict and the minimum identifier. The rule $\RBit$ increases the phase of the node $v$, when $v$ is a player and does not have the minimum identifier at the selected phase. The rule $\RRelay$ updates the identifier variable when $v$ is a relay node.
\vspace*{-0,2cm}
\LinesNumberedHidden 
\begin{algorithm*}
\caption{\AlgoC \label{alg:Color}}
\begin{footnotesize}
\[
\hspace*{-1cm}\vspace*{-0,3cm}
\begin{array}{lcllll}
\RDelta&\sep&(\degr_v\neq \deg(v))&\longrightarrow &\degr_{v}:=\deg(v);\\
\RDeltaP&\sep&(\degr_v= \deg(v)) \wedge (\couleur_{v}>\Delta(v)^{2})&\longrightarrow &\couleur_{v}:=1;\\
\RUp&\sep&(\degr_v= \deg(v)) \wedge (\couleur_{v}\leq\Delta(v)^{2})\wedge\Illicit(v) &\longrightarrow &\AUpdate(v);\\
\RColor&\sep&(\degr_v= \deg(v)) \wedge(\couleur_{v}\leq\Delta(v)^{2})\wedge\neg \Illicit(v) \wedge \Player(v)  \wedge \Loser(v) &\longrightarrow &\Newcolor(v);\\
\RBit&\sep&(\degr_v= \deg(v)) \wedge(\couleur_{v}\leq\Delta(v)^{2})\wedge\neg \Illicit(v) \wedge \Player(v)  \wedge \neg\Loser(v)\wedge \SIncPhase(v,\Oth(v)) &\longrightarrow & \SIncPh(v);\\
\RRelay&\sep&(\degr_v= \deg(v)) \wedge(\couleur_{v}\leq\Delta(v)^{2})\wedge\neg \Illicit(v)  \wedge \RelayC(v) \wedge \RelayUp(v,\PlayerR(v)) &\longrightarrow & \Opt(v,\PlayerR(v),\max);\\

\end{array}
\]
\end{footnotesize}
\end{algorithm*}
\vspace*{-0,3cm}
\begin{theorem}
\label{theo:color}
Algorithm $\AlgoC$ solves the vertex coloration problem at distance two in a silent self-stabi\-li\-zing manner in graph, assuming the state model, and a distributed unfair scheduler. Moreover, if the $n$ node identifiers are in $[1,n^c]$, for some $c\geq 1$, then Algorithm $\AlgoC$ uses $O(\max\{\log \Delta,\log \log n\})$ bits of memory per node. 
\end{theorem}


\subsection{Cleaning a cycle or an impostor-rooted spanning tree}
\label{sub:Freeze}
\vspace{-0,1cm}
The graph $G$ is supposed to be colored up to distance 2, thanks to our previous algorithm. To construct a spanning tree of $G$, each node $v$ maintains a variable $\p_v$ storing the color of $v$'s parent ($\emptyset$ otherwise). The function $\Child(v)$ to return the subset of $v$'s neighbors considered as its children (that is, each such node $u$ has its $p_{u}$ variable equal $v$'s color). Not that the variable parents is managed by the algorithm of spanning tree-construction.

An error is characterized by the presence of inconsistencies between the values of the variables of a node $v$ and those of its neighbors. In the process of a tree-construction, an error occurring at node $v$ may have impact on its descendants. For this reasons, after a node $v$ detects an error, our algorithm cleans $v$ and all of its descendants. The cleaning process is achieved by Algorithm $\AlgoFreeze$, already presented in previous works~\cite{BlinT_DISC13,BlinT17,BlinF15}. Algorithm $\AlgoFreeze$ is run in two cases: cycle detection (thanks to predicate $\ErCycle(v)$, presented in Subsection~\ref{sub:Break}), and impostor leader detection (thanks to predicate $\ErST(v)$, presented in Subsection~\ref{sub:LE}). An impostor leader is a node that (erroneously) believes that it is a root.

 When a node $v$ detects a cycle or an impostor root, $v$ deletes its parent. Simultaneously, $v$ becomes a \emph{frozen} node. Then, every descendant of $v$ becomes frozen. Finally, from the leaves of the spanning tree rooted at $v$, nodes delete their parent and reset all variables that are related to cycle detection or tree-construction. So, this cleaning processe cannot create a livelock. Algorithm \AlgoFreeze\/ is a silent self-stabilizing algorithm using  $O(1)$ bits of memory per node (see Annexe \ref{annexe:Clean}).
\vspace{-0,3cm}
\subsection{Silent self-stabilizing algorithm for cycle detection}
\label{sub:Break}

We present in this subsection a self-stabilizing algorithm to detect cycles (possibly due to initial incorrect configuration) without using the classical method of computing the distance to the root. We first present our solution with the assumption of global identifiers (hence using $O(\log n)$ bits for an $n$-node network), and then using our compact identifier scheme.

\vspace{-0,3cm}
\subsubsection{Self-stabilizing algorithm with identifiers}
\vspace{-0,1cm}
The main idea to detect cycles is to use the uniqueness of the identifiers. We flow the minimum identifier up the tree to the root, then if a node whose identifier is minimum receives its identifier, it can detect a cycle. Similarly, if a node $v$ has two children flowing the same minimum identifier, $v$ can detect a cycle. The main issue to resolve is when the minimum identifier that is propagated to the root does not exists in the network (that is, it results from an erroneous initial state). 

The variable $\minId_v$ stores the minimum identifier collected from the leaves to the root up to node~$v$. We denote by $\del_v$ the minimum identifier obtained by~$v$ during the previous iteration of the protocol (this can be $\emptyset$).
A node $v$ may selects among its children the node $u$ with the smallest propagated identifier stored in $\minId_{u}$, we call this child \emph{kid} returned by the function $\kid(v)$. 

Predicate $\ErCycle(v)$ is the core of our algorithm. Indeed, a node $v$ can detect the presence of a cycle if it has a parent and if \emph{(i)} one of its  child publishes its own identifier, or \emph{(ii)} two of its children publish the same identifier. Let us explain those conditions in more detail. 

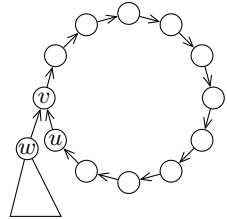
\begin{wrapfigure}[8]{r}{4cm}
\begin{center}
\vspace*{-01cm}
\begin{tikzpicture} [scale=0.45]
	\foreach \x in {0,...,11} 
		\edef\mya{2}
  		\pgfmathparse{int(\mya+\x)}
  		\edef\mya{\pgfmathresult}
		\node [circle,draw,fill=white,scale=0.8] (a\x) at (360*\x/12+90:2.5) {};
	\foreach \x in {1,...,11} 
		\edef\mya{-1}
  		\pgfmathparse{int(\mya+\x)}
  		\edef\mya{\pgfmathresult}
		\path[draw,-angle 45] (a\x)--(a\mya);
	\path[draw,-angle 45] (a0)--(a11);
	\draw (a3) node {\footnotesize $v$};
	\draw (a4) node {\footnotesize $u$};
	\node [circle,draw,fill=white,scale=0.8] (w) at (-3,-1.5) {};
	\draw (w) node {\footnotesize $w$};
	\path[draw,-angle 45] (w)--(a3);
	\path[draw] (w)-- (-3.5,-3.5) -- (-2,-3.5) -- (w);
\end{tikzpicture}
\caption{\footnotesize Spanning structure\label{Fig:Structure}}
\end{center}
\end{wrapfigure}
Let us consider a spanning structure $S$, a node $v\in S$  and let $u$ and $w$ be two of its children. Suppose that $v$ and $u$ belong to a cycle ${\tt C}$, note that, since a node has a single parent, $w$ cannot belong to any cycle (see Figure~\ref{Fig:Structure}).
Let $\widecheck{\minId}$ be the minimum identifier stored by any variable $\minId_{z}$ such that $z$ belongs to $S$. So, $z$ is either in ${\tt C}$, or in the subtree rooted to $w$, denoted by ${\tt T}_w$.

 First, let us consider the case where $\widecheck{\minId}$ is stored in  ${\tt T}_w$. As any node selects the minimum for flowing the $\minId$ upstream, there exists a configuration $\gamma$ where $\minId_{w}=\widecheck{\minId}$, and a configuration $\gamma'>\gamma$ where $\minId_{u}=\widecheck{\minId}$. In $\gamma'$, $v$ can detect an error, due to the uniqueness of identifier, it is not possible for two children of $v$ to share the same value when there is no cycle.

Now, let us suppose that $\widecheck{\minId}$ is in ${\tt C}$, and let $v'$ be the node with the smallest identifier in ${\tt C}$, so $\minId_{v'}=\widecheck{\minId}$ or $\minId_{v'}\neq \widecheck{\minId}$ ($\minId_{v'}\neq \widecheck{\minId}$ means that the identifier $\widecheck{\minId}$ does not exist in ${\tt C}$.) If $\minId_{v'}=\widecheck{\minId}$, as any node selects the minimum for flowing the $\minId$ upstream, there exists a configuration $\gamma$ where $\minId_{u'}=\widecheck{\minId}$ and $u'$ is the child of $v'$ involved in ${\tt C}$, then $v'$ can detect an error. Indeed, due to the uniqueness of identifier, it is not possible that one of its children store its identifier when there is no cycle. The remaining case is when $\minId_{v'}\neq\widecheck{\minId}$. In this case, as any node selects the minimum for flowing the $\minId$ upstream, there exists a configuration $\gamma$ where $\minId_{z}=\widecheck{\minId}$, with $z$ belonging to ${\tt C}$. When a node $v$, its parent and one of its children share the same minimum, they restart the computation of the minimum identifier. For this purpose, they put their own identifier in the $\minId$ variable. To avoid livelock, they also keep track of the previous $\widecheck{\minId}$ in variable $\del_v$. Now $\widecheck{\minId}=\minId_{v'}$, so the system reaches the first case.  Note that  the variable $\del_v$ blocks the livelock but also the perpetual restart of the nodes, as a result of this, a silent algorithm.
Moreover, a node $v$ collects the minimum identifier from the leaves to the root, if $\minId_{v}$ contains an identifier bigger than the identifier of the node $v$, then $v$ detects an error. The same holds, when $v$ has a $\minId_{v}$ smaller than $\minId_{u}$ with $u$ children of $v$, since the minimum is computed between $\minId_{\kid(v)}$ and its own identifier.  

\vspace{-0.4cm}
\begin{footnotesize}
\begin{equation*}
\ErCycle(v)\equiv (\p_v\neq \emptyset) \wedge \Big((\minId_{\kid(v)}=\ID_v) \vee (\exists (u,w)\in \Child(v): \minId_{u}=\minId_{w}) \vee  (\minId_{v}>\ID_{v}) \vee \big( (\minId_{v}\neq \ID_{v})\wedge (\minId_{v}<\minId_{\kid(v)})\big)\Big)
\vspace{-0.6cm}
\end{equation*}
\end{footnotesize}

Our algorithm contains three rules. The first rule $\RMin(v)$ updates the minimum variable $\minId_v$ if the minimum variable $\minId_u$ of a child $u$ is smaller, nevertheless this rule is enabled if and only if the variable $\del_v$ does not contain the minimum $\minId_u$ published by the child. When a node $v$ and its relatives have the same minimum, $v$ declares its intent to restart a minimum identifier computation by erasing its current (and storing it in $\del_v$). The rule $\RStart(v)$ is dedicated to declaring its intent to restart. When all its neighbors have the same intent, the node can restart (see rule $\RID(v)$).
\LinesNumberedHidden 
\begin{small}
\vspace{-0,3cm}
\begin{algorithm*}
\caption{ Algorithm \AlgoBreakR\/ For node $v$ with  $\neg\ErCycle(v)$}
\[
\hspace*{-1cm}
\begin{array}{lcllll}
\RMin&\sep& (\minId_{v}>\minId_{\kid(v)}) \wedge (\del_{v}\neq \minId_{\kid(v)})&\longrightarrow \minId_v:=\minId_{\kid(v)};\\
\RStart&\sep&  (\minId_{\p_{v}}=\minId_{v}=\minId_{\kid(v)})\wedge (\del_v\neq \minId_{v})&\longrightarrow \del_v:=\minId_{v};\\
\RID&\sep& (\del_{\p_{v}}=\del_{v}=\del_{\kid(v)}=\minId_{v})\wedge (\minId_{v}\neq \ID_{v})&\longrightarrow \minId_{v}:=\ID_{v};\\
\end{array}
\]
\vspace{-0,3cm}
\end{algorithm*}
\end{small}

\vspace{-0,4cm}
\begin{theorem}
\label{theo:CycleR}
Algorithm \AlgoBreakR\/ solves the  detection of cycle in $n$-node graph  in a silent self-stabi\-li\-zing manner, assuming the state model, and a distributed unfair scheduler. Moreover, if the $n$ node identifiers are in $[1,n^c]$, for some $c\geq 1$, then algorithm \AlgoBreakR\/  uses $O(\log n)$ bits of memory per node.
\end{theorem}
\vspace{-0,6cm}
\subsubsection{Talkative self-stabilizing cycle detection with compact identifiers}
\vspace{-0,1cm}
We  refine  algorithm \AlgoBreakR\/  to make use of compact identifiers (of size $O(\log \log n)$ instead of global identifiers (of size $O(\log n)$). With compact identifiers, the main problem is the following: two nodes $u$ and $v$ can deduce that $\PactID_u=_{c}\PactID_u$ if and only if they have observed $\PactID_u\simeq_{c}\PactID_v$ during every phase $i$, with $1\leq i\leq \SB_{v}$. 
 A node $v$ selects the minimum compact identifier stored in variable $\minId$ in its neighborhood (including itself). 
 If in a previous configuration one of its children had presented $v$ a compact identifier smaller than its own, $v$ became passive (Variable $\Active_v=false$), and remained active otherwise (Variable $\Active_v=true$).
 Only active nodes  can continue to increase their phase. Moreover, a node increases its phase if and only if its parent and one of its children $u$ has the same information, namely $\PactID_u\simeq_{c}\PactID_v\simeq_{c}\PactID_{\p_v}$.
 Note that, in a spanning tree several nodes may  not increase their phases, for example leaves which by definition have no child, this does not cause a problem. Let us explain, let $v$ be the node with the smallest identifier involved in a cycle  and let suppose that $v$ has two children, one $u$ involved in the cycle and the child $w$ no. In some configuration the node $w$ has not able to increase its phase, but the node $u$ will reach the same phase of the active node $v$, so $v$ increase its phase, and  the system reaches a configuration where $\PactID_u=_{c}\PactID_v$ so $v$ detects an error of cycle.  The variable $\del_v$ combined to this compact identifier use allow us to maintain a silent algorithm.

Predicate $\ErCycle$ now takes into account the error(s) related to compact identifiers management. It is important to note that the cycle breaking algorithm does not manage phase differences. Indeed, a node $v$ with a phase bigger than the phase of one of its children $u$ takes the $\minId_u$, if and only if its phase its bigger than two or if no child has its same compact identifier.
The  $\minId_{v}$ variable is be compared using lexicographic order by rule $\RMin$.
%
%
Tthe modifications to algorithm $\AlgoBreakR$ are  minor. We add only one rule to increases the phase: $\RInc$. Only a passive node can restart.
Remark that now the $\minId$ variable uses $O(\log \log n)$ bits. As the $\p$ variables stores a color, we obtain a memory requirement of $O(\max\{\log \Delta, \log \log n\})$ bits per node. 

\vspace{-0,2cm}
\begin{theorem}
\label{theo:CCycle}
Algorithm \CAlgoBreakG\/ solves the  detection of cycle in arbitrary $n$-node graph in a silent self-stabi\-li\-zing manner, assuming the state model, and a distributed unfair scheduler. Moreover, if the $n$ node identifiers are in $[1,n^c]$, for some $c\geq 1$, then algorithm \CAlgoBreakG\/  uses $O(\max\{\log \Delta, \log \log n\})$ bits of memory per node.
\end{theorem}


\vspace{-0,5cm}
\subsection{Spanning tree-construction without distance to the root maintenance}
\label{sub:LE}
\vspace{-0,1cm}
Our approach for self-stabilizing leader election is to construct a spanning tree whose root is to be the elected leader. Two main obstacles to self-stabilizing tree-construction are the possibility of an arbitrary initial configuration containing one or more cycles, or the presence of one or more impostor-rooted spanning trees. We already explained how the cycle detection and cleaning process takes place, so we focus in this section on \emph{cycleless} configurations. 

The main idea is to mimics the \emph{fragments} approach introduced by Gallager \emph{et al.}~\cite{GallagerHS83}. In an ideal situation, at the beginning each node is a fragment, each fragment merges with a neighbor fragment holding a bigger root signature, and at the end remains only one fragment, rooted in the root with the biggest signature (that is, the root with maximum degree, maximum color, and maximum global identifier). To maintain a spanning structure, the neighbors that become relatives (that is, parents or children) remain relatives thereafter.
Note that the relationship may evolve through time (that is, a parent can become a child and vice versa). So our algorithm maintains that as an invariant, given by Lemma~\ref{lem:invariant}.

Indeed, when two fragments merge, the one with the root with smaller signature $F_{1}$ and the other one with a root with bigger signature $F_{2}$, the root of $F_{1}$ is re-rooted toward its descendants until reaching the node that identified $F_{2}$. This approach permits to construct an acyclic spanning structure, without having to maintain distance information. 
The variable $\R_v$ stores the signature relative to the root (that is, its degree, its color, and its identifier). Note that, the comparison between two $\R$ is done using lexical ordering. Moreover, the variable $\new_v$ stores the color of the neighbor $w$ of $v$ leading to the a node $u$ with $\R_u>\R_v$ if there exists such a node, and $\emptyset$ otherwise.
The function $\Cand(v)$ returns the color of the neighbor of $v$ with the maximum root.

Let us now give more details about our algorithm (presented in Algorithm~\ref{Algo:ST}). If a root $v$ has a neighbor $u$ with  $\R_u>\R_v$, then $v$ chooses $u$ as its parent (see rule $\RMerge$ ). If a node $v$ (not a root) has a neighbor $u$ with $\R_u>\R_v$, it stores its neighbor's color in Variable $\new_v$, and updates its $\R_v$ to $\R_u$. Yet, it does not change its parent. This behavior creates a path (thanks to Variable $\new$) between a root $r$ of a sub spanning tree ${\tt T}_r$ and a node contained in an other sub spanning tree ${\tt T}_{r'}$ rooted in $r'$, with $\R_{r'}>\R_r$ (see rule $\RPath$). The subtree ${\tt T}_r$ is then re-rooted toward a node aware of a root with a bigger signature $u$. Now, when $v\in {\tt T}_r$'s neighbor  $u$ becomes root, it takes $u$ as a parent (see rules $\RReRoot$ and $\RDel$). Finally, the descendants of the re-rooted root update their root variables (see rule $\RUpdate$).
\vspace*{-0.1cm}
\LinesNumberedHidden 
\begin{algorithm*}

\caption{ Algorithm $\AlgoST$ \label{Algo:ST}}
\[
\hspace*{-1cm}
\begin{footnotesize}
\vspace*{-0.5cm}
\begin{array}{lcllll}
\RDel&\sep& (\p_{\p_v}=\couleur_v) &\longrightarrow \p_{v}:=\emptyset;\\
\RUpdate&\sep&  (\p_{\p_v}\neq\couleur_v) \wedge  (\Cand({v})\neq \emptyset) \wedge (\R_{v}<\R_{\Cand(v)})  \wedge (\p_{v}=\Cand(v)) \wedge (\new_v=\emptyset)&\longrightarrow \R_{v}:=\R_{\Cand} ;\\
\RPath&\sep& (\p_{\p_v}\neq\couleur_v) \wedge  (\Cand(v)\neq \emptyset)\wedge(\R_{v}<\R_{\Cand(v)}) \wedge  (\p_{v}\not\in\{\emptyset,\Cand(v)\}) \wedge (\new_v=\emptyset)&\longrightarrow (\R_v,\new_v):=(\R_{\Cand(v)},\Cand(v)) ;\\
\RMerge&\sep& (\p_{v}=\emptyset) \wedge  (\Cand(v)\neq \emptyset)\wedge(\R_{v}<\R_{\Cand(v)})    \wedge (\new_{\Cand(v)}=\emptyset)&\longrightarrow (\p_{v},\R_v):=(\Cand(v),\R_{\Cand}) ;\\
\RReRoot&\sep&(\p_{v}=\emptyset) \wedge  (\Cand(v)\neq \emptyset)\wedge (\R_{v}=\R_{\Cand(v)}) \wedge (\new_v\neq\emptyset) &\longrightarrow (\p_v,\new_v):=(\new_v,\emptyset) ;\\
\end{array}
\end{footnotesize}
\]

\end{algorithm*}
\vspace*{-0.2cm}
The predicate $\ErST(v)$ captures trivial errors and impostor-root errors for the construction of the spanning tree, these errors are formalized in predicate $\ErST(v)$ formalized in the appendix. Note that predicate $\ErST(v)$ is used in $\AlgoFreeze$ only (and not in $\AlgoST$) as these errors are never created by $\AlgoST$ and $\AlgoFreeze$ has higher priority than $\AlgoFreeze$ (see Section~\ref{sub:CorST}). 
\vspace*{-0,2cm}
\label{sub:CorST}
\begin{theorem}
\label{theo:ST}
Algorithm \AlgoST\/ solves the spanning tree-construction problem in a silent self-stabilizing manner in any $n$-node graph, assuming the absence of spanning cycle, the state model, and a distributed unfair scheduler, using $O(\log n)$ bits of memory per node.
\end{theorem}

\vspace*{-0.5cm}
\subsubsection{Spanning tree-construction with compact identifiers}
\vspace*{-0.2cm}
We adapt our algorithm \AlgoST\/ to use compact identifiers and obtain Algorithm \AlgoCST. 
It is simple to compare two compact identifiers  when the nodes are neighbors. Yet, along the algorithm execution, some nodes become non-root, and therefore the remaining root of fragments can be far away, separated by non-root nodes. 
To enable multi-hop comparison, we use a broadcasting and convergecast wave on a spanning structure. Let $v$ a node that wants to broadcast its compact identifier. We add an variable $\chek$ to our previous algorithm. This variable  checks whether every descendant or neighbor shares the same compact identifier at the same phase before proceeding to the convergecast. More precisely, a node $u$ must checks if every neighbors $w$ has $\PactID_u\simeq_{c}\PactID_w$, and if every child has $\chek_{v}=true$. If so, it sets its variable $\chek_{v}=true$, and the process goes on until node $v$. As a consequence, $v$ increases or restarts its phase and assigns $false$ to $\chek$. \vspace*{-0.2cm}
\begin{theorem}
\label{theo:CST}
Algorithm \AlgoCST\/ solves the spanning tree-construction problem  in a talkative self-stabilizing manner in any $n$-node graph, assuming the absence of spanning cycle, the state model, and a distributed unfair scheduler, in $O(\max\{\log \Delta, \log \log n\})$ bits of memory per node.
\end{theorem}
\vspace*{-0.2cm}


\vspace*{-0.4cm}
\section{Self-stabilizing leader election}
\label{sec:Cor}
\vspace*{-0.3cm}
We now present the final assembly of tools we developed to obtain a self-stabilizing leader election algorithm. 
We add to Algorithm  $\AlgoCST$ an extra variable $\ell$ that is mainntained as follows: if a node $v$ has no parent, then $\ell_v=true$, otherwise, $\ell_v=false$. Variable $\ell_v$ is meant to be the output of the leader election process.

Proof sketch of Theorem \ref{theo:CompoC}: Our self-stabilizing leader election algorithm results from combining severals algorithms. As already explained, a higher priority algorithm resets all the variables used by lesser priority algorithms. Moreover, lesser priority algorithm do not modify the variables of the higher priority algorithms. Algorithms are prioritized as follows: 
$\AlgoC$,  $\AlgoFreeze$, $\CAlgoBreakG$ and  $\AlgoCST$.
Only the algorithm $\AlgoCST$ is talkative,  we first proof that the number of activations of rules of algorithm $\AlgoCST$ are bounded if there exist nodes enabled by $\AlgoC$, $\AlgoFreeze$ or $\CAlgoBreakG$. So we already proof the convergence of algorithms $\AlgoC$, $\AlgoFreeze$ and $\CAlgoBreakG$. Thanks to Theorem~\ref{theo:CST}, we obtain a spanning tree rooted in the node with the maximum degree, maximum color, and maximum identifier. As a consequence, only the root $r$ has $\ell_r=true$ and every other node $v\in V\setminus\{r\}$ has $\ell_v=false$.

\newpage
\bibliographystyle{plain}	
\bibliography{BibPointeurs}

\newpage
\appendix
\makeatother
\section{Appendix}

\subsection{Compact memory using identifiers: Predicates}
Node $v$ can trivially detect an error (predicate $\ErT(v)$) whenever its compact identifier does not match its global identifier, or its phase is greater than $\SB_v$.
\begin{equation}
\label{eq:ErT}
\ErT(v)\equiv  \big[\PactID_v\neq(\Bit(1,\ID_v),\Phase_v,\Bit(\Phase_v,\ID_v))\big] \vee (\Phase_v>\SB_{v}) 
\end{equation}

Moreover, in normal operation, the phases of neighboring nodes must be close enough: a node's phase may not be more than $1$ ahead or behind any of its neighbors; also a node may not have a neighbor ahead and another behind. 

Predicate $\SErB(v,S)$ captures these conditions, where $S(v)$ denotes a subset of neighbors of $v$. The set $S$ should be understood as an input provided by an upper layer algorithm. 
\begin{equation}
\label{eq:SErB}
\begin{split}
\SErB(v,S)\equiv  \
\Big(\exists u,w\in S(v):\big(\Phase_{u}>\Phase_{v}+1\big) \vee \big(\Phase_{u}<\Phase_{v}-1 \big)\vee 
(|\Phase_{u}-\Phase_{w}|=2) \Big)
\end{split}
\end{equation}

In a talkative process, node identifiers are published (though compact identifiers) infinitely often. So, when node $v$ and all its active neighbors have reached the maximum phase (\emph{i.e.} $\Phase_{v}=\SB_{v}$), $v$ goes back to phase $1$. Then, if $v$ has $\Phase_{v}=\SB_{v}$ and an active neighbor $u$ has $\Phase_{u}=1$, it is not an error. But if $v$ has $\Phase_{v}=1$, one active neighbor $u$ has $\Phase_{u}=\SB_{v}$, and another active neighbor $w$ has $\Phase_{w}>1$, then an error is detected.
\begin{equation}
\label{eq:TErB}
\begin{split}
\TErB(v,S)\equiv \Big(\exists u,w\in S(v): 
 \big[ (1<\Phase_v<\SB_{v}) \wedge ((\Phase_{u}>\Phase_{v}+1\big) \vee \big(\Phase_{u}<\Phase_{v}-1))\big] \vee\\
\big[ (\Phase_{v}=\SB_{v}) \wedge \big((\Phase_{u}>1) \vee (\Phase_{u}<\SB_{v}-1) \vee ((\Phase_{u}=\Phase_{v}-1)\wedge (\Phase_{w}=1))\big)\big]\vee \\
\big[ (\Phase_{v}=1) \wedge   \big((\Phase_{u}>2) \vee (\Phase_{u}<\SB_{v}) \vee ((\Phase_{u}=\SB_{v})\wedge (\Phase_{w}=2))\big)\big]
\Big)
 \end{split}
\end{equation}

If $v$ detects an error through $\ErT(v)$, $\SErB(v,S)$ or $\TErB(v,S)$, it resets its compact identifier to its first phase value:
\begin{equation}
\ResetCID(v):\PactID_v:=(\SB_{v},\Phase_v,\Bp_v)=(\Bit(1,\ID_v),1,\Bit(1,\ID_v))
\label{Com:ResetCID}
\end{equation}
This may trigger similar actions at neighbors in $S$, so that all such errors eventually disappear.

The compact identifier of $u$ is smaller (respectively greater) than the compact identifier of $v$, if the most significant bit-position of $u$ is smaller (respectively greater) than the  most significant bit-position of $v$, or if the most significant bit-position of $u$ is equal to the  most significant bit-position of $v$, $u$ and $v$ are in the same phase, and the bit-position of $u$ is smaller (respectively greater) than the bit-position of $v$:
\begin{equation}
\begin{split}
 \PactID_u^{\,i}<_{c}\PactID_v^{\,i}\equiv(\SB_{u}<\SB_{v}) \vee   \big((\SB_{v}=\SB_{u}) \wedge (\Bp_{u}<\Bp_{v})\big) 
\label{P:Inf}
\end{split}
\end{equation}
%
When two nodes $u$ and $v$ have the same most significant bit-position and the same bit position at phase $i<\SB_{v}$, they are possibly equal with respect to compact identifiers (denoted by $\simeq_c$).
\begin{equation}
\begin{split}
\PactID_u^{\,i}\simeq_{c}\PactID_v^{\,i}\equiv    (i<\SB_v) \wedge\big((\SB_{v}=\SB_{u}) \wedge  (\Bp_{u}=\Bp_{v})\big) 
\label{P:Meq}
\end{split}
\end{equation}

Finally, two nodes $u$ and $v$ have the same compact identifier (denoted by $=_c$) if their phase reaches the size of the binary representation of the identifier of the two nodes, and their last bit-position is the same.
\begin{equation}
\begin{split}
\PactID_u^{\,i}=_{c}\PactID_v^{\,i}\equiv (i=\SB_{v}=\SB_{u}) \wedge (\Bp_{u}=\Bp_{v})
\label{P:Eq}
\end{split}
\end{equation}

Predicate $\SIncPhase(v)$ is true if for every node $u$ in $S(v)$, either $\PactID_u^{\,i}\simeq_{c}\PactID_v^{\,i}$, or $\Phase_u=\Phase_v+1$. 
\begin{equation}
\begin{split}
\SIncPhase(v,S)\equiv \forall u\in S(v): (\PactID_u^{\,i}\simeq_{c}\PactID_v^{\,i}) \vee (\Phase_u=\Phase_v+1)
\label{P:IncPhase}
\end{split}
\end{equation}

Similarly, $\TIncPhase(v,S)$ is true if for every node $u$ in $S(v)$, either $\PactID_u^{\,i}=_{c}\PactID_v^{\,i}$, or $\Phase_u=1$. 
\begin{equation}
\begin{split}
\TIncPhase(v,S)\equiv \SIncPhase(v,S) \vee \forall u\in S(v): (\Phase_v=\SB_v)\wedge \big((\PactID_u^{\,i}=_{c}\PactID_v^{\,i}) \vee  (\Phase_u=1)\big)
\label{P:IncPhase}
\end{split}
\end{equation}

When $\TIncPhase(v,S)$ or $\SIncPhase(v)$  is true, $v$ may increase its phase:
\begin{equation}
\TIncPh(v): \PactID_v:=\left\{
\begin{array}{ll}
(\SB_{v},\Phase_{v}+1,\Bit(\Phase_{v}+1,\ID_v)) &\text{ if } \Phase_v<\SB_v\\
(\SB_{v},1,\Bit(1,\ID_v)) &\text{ if } \Phase_v=\SB_v\\
\end{array}
\right.
\label{C:IncPh}
\end{equation}

In some case, we need to compute the minimum or the maximum on compact identifiers. Let $f$ denote a function that is either minimum or maximum. Let us denote by  $\CB(v,S,f)$ the minimum or the maximum most significant bit of nodes in $S(v)$.
\begin{equation}
\begin{split}
\CB(v,S,f)= f\{\SB_{w}:w\in S(v)\} 
\label{P:}
\end{split}
\end{equation}
To compare compact identifiers, one must always refer to the same phase; we always consider the minimum phase for nodes in $S(v)$.
\begin{equation}
\begin{split}
\CPh(v,S,f)= \min\{\Phase_{w}:w\in S(v), \SB_{w}=\CB(w,S,f)\} 
\label{P:}
\end{split}
\end{equation}
Finally, we compute the minimum or the maximum bit position.
\begin{equation}
\begin{split}
\CBp(v,S,f)= f\{\Bp_{w}:w\in S(v), \Phase_{w}=\CPh(w,S,f)\}
\label{P:}
\end{split}
\end{equation}
Predicate $\MinCID(v,S)$ checks if $\PactID_{v}$ is equal to the minimum among nodes in $S(v)$:
\begin{equation}
\begin{split}
\MinCID(v,S)\equiv \big(\PactID_{v}= (\CB(v,S,\min),\CPh(v,S,\min),\CBp(v,S,\min))\big)
\label{P:}
\end{split}
\end{equation}
The predicate $\MaxCID(v,S)$ does the same for the maximum:
\begin{equation}
\begin{split}
\MaxCID(v,S)\equiv \big(\PactID_{v}= (\CB(v,S,\max),\CPh(v,S,\max),\CBp(v,S,\max))\big)
\label{P:}
\end{split}
\end{equation}

Node $v$ may use $\Opt$ to assign its local variables the minimum (or maximum) compact identifier in $S(v)$.
\begin{equation}
\Opt(v,S,f):\left\{ 
\begin{array}{l} 
 \SB_{v}:= \CB(v,S,f)\\
 \Phase_{v}:= \CPh(v,S,f)\\
 \Bp_{v}:= \CBp(v,S,f)\\
\end{array} \right.
\label{C:Verif}
\end{equation}



\subsection{Silent Self-stabilizing Distance-2 Coloring: Predicates and Correctness}
\subsubsection{Predicates}

Note that all rules are exclusive, because a node $v$ cannot be both $\Player(v)$ and $\RelayC(v)$.
Let us now describe the functions, predicates and actions use by  algorithm \AlgoC.
Remember that $N[v]=N(v)\cup\{v\}$. 
Function $\Delta(v)$ returns the maximum degree between $v$ and its neighbors, and is used to define the range $[1,\Delta(v)^{2}+1]$ of authorized colors  for a node $v$:
\begin{equation}
\Delta(v)=\max\{\degr_u:u\in N[v]\} 
\label{eq:Delta}
\end{equation}

The function $\MinCf(v)$ returns the minimum color in conflict at distance one and two  :
\begin{equation}
\MinCf(v):=\min\{\couleur_{u}: u,w\in N[v] \wedge (u\neq w) \wedge (\couleur_{u}=\couleur_{w})\}
\label{F:MinCf}
\end{equation}

The function $\MaxC(v)$ returns the   maximum color used at distance one :
\begin{equation}
\MaxC(v):=\max\big\{\couleur_{u}: u\in N[v] \big\}
\label{F:MaxC}
\end{equation}

The predicate $\Illicit(v)$ is true if $v$ has not yet set the right value for either $\MinCf(v)$ or $\MaxC(v)$. Moreover, this predicate checks if $v$'s compact identifier matches its global identifier (see predicate \ref{eq:ErT}: $\ErT(v)$) and if the phases of the subset of $v$'s neighbors $\Oth(v)$ are coherent with $v$ (see predicate \ref{eq:SErB}: $\SErB(v,\Oth(v)$).

\begin{equation}
\begin{split}
\Illicit(v)\equiv (\conflit_{v}\neq\MinCf(v))\vee (\Mc_{v}\neq\MaxC(v)) \vee  (\ErT(v) \vee \SErB(v,\Oth(v))
\label{P:Illicit}
\end{split}
\end{equation}

The predicate $\Player(v)$ is true if $v$ has the minimum color in conflict (announced by its neighbors or by itself). Observe that the conflict may be at distance one or two:
\begin{equation}
\begin{split}
\Player(v)\equiv (\couleur_{v} =\min \{\conflit_{u},u\in N[v]\}) 
\label{P:Player}
\end{split}
\end{equation}

The predicate $\RelayC(v)$ is true if $v$ does not have the minimum color in conflict, and at least two of its neighbors have the minimum color in conflict:
\begin{equation}
\begin{split}
\RelayC(v)\equiv (\conflit_{v}\neq\couleur_{v}) \wedge ( \conflit_{v}=\min \{\conflit_{u},u\in N[v]\})\wedge (\exists  u,w \in N(v),(\conflit_v=\couleur_u)\wedge (\conflit_v=\couleur_w)) 
\label{P:Relay}
\end{split}
\end{equation}

\begin{figure}[!th]
\begin{center}

\begin{tikzpicture}
   \tikzstyle{VertexStyle}=[shape        = circle,
                            fill  = green!90!white,
                            minimum size = 18pt,
                            draw]

        \Vertex[x=4, y=0,L={\tiny $C,1$}]{C}
	 \node at (4.4,0.5) {\tiny $2,2$};
	 \Vertex[x=8, y=0,L={\tiny $E,1$}]{E}
	 \node at (8.4,0.5) {\tiny $\bot,2$};
	  
	\tikzset{VertexStyle/.append style={fill  = blue!80!white}}
	\Vertex[x=0, y=0,L={\tiny $A,3$}]{A}
	\node at (0.4,0.5) {\tiny $\bot,3$};

	\tikzset{VertexStyle/.append style={fill  = red!80!white}}
	 \Vertex[x=2, y=0,L={\tiny $B,2$}]{B} 
       \node at (2.4,0.5) {\tiny $\bot,3$};
       \Vertex[x=6, y=0,L={\tiny $D,2$}]{D} 
       \node at (6.4,0.5) {\tiny $1,2$};
       
	\Edge(A)(B)
	\Edge(B)(C)
	\Edge(C)(D)
	\Edge(D)(E)
\end{tikzpicture}
\caption{The pair in the node are the identifier of the node and the color of the node. The pair outside the node are the minimum color in conflict (or $\bot$ if none) and the maximum color used. Node $D$ is a relay for node $C$ and $E$ because $\conflit_D=1$, the color of $C$ and $V$. Similarly, Node $C$ is a relay for nodes $B$ and $D$. }
\label{default}
\end{center}
\end{figure}
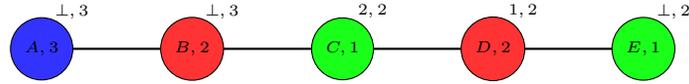

The function $\PlayerR(v)$ returns the subset of $v$'s neighbors that have the minimum color in conflict, when $v$ is a relay node:
\begin{equation}
\PlayerR(v):=\{u: u\in N(v) \wedge \couleur_{u}=\conflit_{v}\} 
\label{F:PlayerR}
\end{equation}
The function $\Oth(v)$ returns the subset of $v$'s neighbors that are in conflict with $v$ at distance one, or the set of relay nodes for the conflict at distance two, when $v$ has $\Player(v)$ equal to $true$:
\begin{equation}
\Oth(v):=\{u: u\in N(v) \wedge \couleur_{u}=\couleur_{v}\} \cup \{u: u\in N(v) \wedge \conflit_{u}=\couleur_{v}\}
\label{F:OtherP}
\end{equation}
The predicate $\Loser(v)$ is true whenever a competing player of  $v$ has a greater bit position at the same phase. A node whose identifier is maximum among competitors does not change its color, but any loosing competitor does.
\begin{equation}
\begin{split}
\Loser(v)\equiv 
\exists u\in \Oth(v):\PactID_v^{\,i}<_{c}\PactID_u^{\,i} 
\label{P:Loser}
\end{split}
\end{equation}


The predicate $\RelayUp(v)$ is true if a relay node is not according to its player neighbors, like we decide to change the color of the node with the minimum identifier the relay node stores the maximum compact identifier of its player neighbors:

\begin{equation}
\begin{split}
\RelayUp(v,\PlayerR(v))\equiv \PactID_{v}\neq_c \MaxCID(v,\PlayerR(v))
\end{split}
\end{equation}



The action $\AUpdate(v)$ updates the variables $\conflit_v,\Mc_v$ and resets the variables relatives to the identifier (see command \ResetCID(v) in equation \ref{Com:ResetCID} ).
\begin{equation}
\AUpdate(v):
\conflit_{v}:=\MinCf(v); \Mc_{v}:=\MaxC(v); \ResetCID(v);
\label{C:Verif}
\end{equation}

When a node change its color, it takes the maximum color at distance one and two plus one modulo $\Delta(v)^{2}+1$, and then add one to assign colors in the range $[1,\dots,\Delta(v)^{2}+1]$.
\begin{equation}
\begin{split}
\Newcolor(v): \couleur_{v}:=\big( (\max\{\MaxC_{u}: u\in N[v]\}+1)\mod \Delta(v)^{2}+1\big) +1; \\
\end{split}
\label{C:Verif}
\end{equation}

\subsubsection{Correctness }
In the details of lemmas that are presented in the sequel, we use predicates on configurations. These predicates are mean to be intermediate attractors towards a legitimate configuration (\emph{i.e.}, a configuration with a unique  leader). To establish that those predicates are indeed attractors, we use potential functions~\cite{T09bc}, that is, functions that map configurations to non-negative integers, and that strictly decrease after any algorithm step is executed.

To avoid additional notations, we use sets of configurations to define predicates; the predicate should then be understood as the characteristic function of the set (that returns \emph{true} if the configuration is in the set, and \emph{false} otherwise). 
\bigskip

\begin{lemma}
Using a range of $[1,\Delta(v)^{2}+1]$ for colors at node $v$ is sufficient to enable distance-2 coloring of the graph.
\label{lem:Delta}
\end{lemma}
\begin{proof}
  In the worst case, all neighbors at distance one and two of $v$ have different colors. Now, $v$ has at most $\Delta(v)$ neighbors at distance one, each having $\Delta(v)-1$ other neighbors than $v$. In total, $v$ has at most $\Delta(v)^2-\Delta(v)$ neighbors at distance up to two, each having a distinct color. Using a range of $[1,\Delta(v)^{2}+1]$ for $v$'s color leaves at least $\Delta(v)+1$ available colors for node $v$.
\end{proof}

Let $\lambda:\Gamma\times V \rightarrow \mathbb{N}$ be the following function: 
$$\lambda(\gamma,v)=
\left\{
  \begin{array}{ll}
   2 &\text{if } \degr_v\neq \deg(v) \text{in} \gamma \\
   1 &\text{if } (\degr_v= \deg(v) \text{in} \gamma) \wedge \couleur_{v}(\gamma)>\Delta(v)^{2}+1\\
   0  &\text{otherwise} \\
  \end{array}
\right.
$$
Let $\Lambda:\Gamma \rightarrow \mathbb{N}$ be the following potential function: 
$$\Lambda(\gamma)=\sum_{v\in V}\lambda(\gamma,v)$$
Let $\tau:\Gamma\times V \rightarrow \mathbb{N}$ be the following function: 
$$\tau(\gamma,v)=
\left\{
  \begin{array}{ll}
   \Delta(v)^{2}+1+\couleur_{v}(\gamma_{0})-\couleur_{v}(\gamma) &\text{if } \couleur_{v}(\gamma)>\couleur_{v}(\gamma_{0})\\
   \couleur_{v}(\gamma_{0})-\couleur_{v}(\gamma)  &\text{otherwise} \\
  \end{array}
\right.
$$
Where $\couleur_{v}(\gamma_{0})$ is the color of $v$ in configuration $\gamma_{0}$, $\gamma_0$ being defined as the configuration where $\Lambda(\gamma_{0},v)$ reaches zero. Also, $\couleur_{v}(\gamma)$ is the color of $v$ in configuration $\gamma$.

Let ${\mathcal C}:\Gamma \rightarrow \mathbb{N}$ be the following potential function: 
$${\mathcal C}(\gamma)=(\kappa_0(\gamma),\kappa_1(\gamma),\dots,\kappa_{\Delta^{2}}(\gamma))$$
where $\kappa_i(\gamma)= |\{ v\in V :\tau(\gamma,v)=i\}|$.
The comparison between two configurations ${\mathcal C}(\gamma)$ and ${\mathcal C}(\gamma')$ is done using lexical ordering.  
We denote by $\gamma'$ the configuration after activation of (a subset of) the nodes in $A_\kappa(\gamma)$ where $A_\kappa(\gamma)$ denotes the enabled  nodes in $\gamma$ due to rule $\RColor$.
We can now prove the following result: ${\mathcal C}(\gamma')<{\mathcal C}(\gamma)$ for every configuration $\gamma$ where $A_\kappa(\gamma)$ is not empty. 

\begin{lemma}
 ${\mathcal C}(\gamma')<{\mathcal C}(\gamma)$ for every configuration $\gamma$ such that $A_\kappa(\gamma)$ is not empty. 
 \label{lem:color}
\end{lemma}

\begin{proof}
We consider a node $v\in A_\kappa(\gamma)$. After executing rule $\RColor$, $v$ takes a color $\couleur_v(\gamma')=\big( (\max\{\MaxC_{u}: u\in N[v]\}+1)\mod \Delta(v)^{2}+1\big)+1$. As a consequence $\tau(\gamma,v)$ decreases by one, 
so ${\mathcal C}(\gamma')<{\mathcal C}(\gamma)$.
\end{proof}

Let $\psi: \Gamma\times V \rightarrow \mathbb{N}$ be the function defined by:
$$\psi(\gamma,v)=
\left\{
  \begin{array}{ll}
   n &\text{if }\Illicit(v) \vee (\SB_{v},\Phase_{v},\Bp_{v}) \; \mbox{is true}\\
  2\log n -\Phase_v & \text{if } \Player(v) \mbox{ is true}\\
  2 & \text{if } \RelayC(v) \wedge \neg\RelayUp(v,\PlayerR(v))\mbox{ is true}\\
    1 & \text{if } \RelayC(v) \wedge  \RelayUp(v,\PlayerR(v))\mbox{ is true}\\
     0 &\text{otherwise} \\
  \end{array}
\right.
$$

Let $\Psi: \Gamma \rightarrow\mathbb{N}$ be the potential function defined by:
$$\Psi(\gamma)=\sum_{v\in V} \psi(\gamma,v)$$ 

Let $\Phi: \Gamma\rightarrow\mathbb{N}$ be the potential function defined by:
$$\Phi(\gamma)=(\Lambda(\gamma),{\mathcal C}(\gamma),\Psi(\gamma)).$$ 

The comparison between two configurations $\Phi(\gamma)$ and $\Phi(\gamma')$ is by using lexical order.  
We denote by $A(\gamma)$ the (subset of) enabled nodes (for any rule of our algorithm) in configuration~$\gamma$.
Note that the algorithm is stabilized when every node is neither a player nor a relay, that is the nodes have no conflict at distance one and two, when $\Psi(\gamma)=0$. We define 
$$\Gamma_{\cal C}=\{\gamma\in \Gamma:\Phi(\gamma)=0\}$$

%
\begin{lemma}
$true \triangleright \Gamma_{\cal C}$ and $\Gamma_{\cal C}$ is closed.
\label{lem:DecColor} 
\end{lemma}

\begin{proof}
The function $\Lambda(\gamma)$ decreases by any execution of rules $\RDelta$ and $\RDeltaP$. Remark that $\deg(v)$ is considered a non corruptible local information, so once $v$ has executed $\RDelta$, this rule remains disabled afterwards. Moreover, $\RColor$ maintains the value of the color inferior (or equal) to $\Delta(v)^{2}+1$, and other rules modifying the color maintain this invariant. Hence, if the scheduler activates rules $\RDelta$ or $\RDeltaP$, we obtain $\Lambda(\gamma')< \Lambda(\gamma)$, otherwise if other rules are activated, then $\Lambda(\gamma')= \Lambda(\gamma)$.
We already saw that, when the scheduler activates a node $v$ for rule $\RColor$,  we obtain ${\mathcal C}(\gamma')<{\mathcal C}(\gamma)$. 
Overall, if the scheduler activates rules $\RDelta$, $\RDeltaP$, or $\RColor$ we obtain $\Phi(\gamma')<\Phi(\gamma)$. We now consider the cases where the scheduler activates other rules.

First, we focus on rule $\RUp$. Let us consider $A'$, the set of nodes enabled for this rule, and a node $v$ such that $v\in A'$. Then, $v$ has $\Illicit(v)=true$ (see predicate \ref{P:Illicit}), or $(\Player(v)  \wedge (\ErT(v) \vee \SErB(v,\Oth(v))) )$ is $true$. 
 If $v$ has $(\conflit_{v}\neq\MinCf(v))\vee (\Mc_{v}\neq\MaxC(v))$ in $\gamma$, then after activation of $v$, we obtain $\conflit_{v}=\MinCf(v)$ and $\Mc_{v}=\MaxC(v)$ in $\gamma'$ because $\MinCf(v)$ and $\MaxC(v)$ depend only on the color of the neighbors of $v$ (see Function~\ref{F:MinCf} and~\ref{F:MaxC}).
The same argument applies for $ (\ErT(v) \vee \SErB(v,\Oth(v)))$, because $\PactID_{v}$ is computed only with the identifier of $v$. So, after execution of $\RUp$ by $v$, we obtain $\psi(\gamma',v)<\psi(\gamma,v)$. Remark that, if the color of the neighbors of $v$ does not change, rule $\RUp$ remains disabled. Now, if the color changes, $\Phi(\gamma)$ still decreases thanks to Lemma~\ref{lem:color}. 
 
Let us consider now a configuration where the rule $\RUp$ is disabled for every node.
Rule $\RBit$ increases the phase of a player node, so after activation of this rule we obtain  $\psi(\gamma',v)<\psi(\gamma,v)$. Executing rule $\RRelay$ decreases also $\psi(\gamma',v)$ due to $\RelayUp(v,\PlayerR(v))$, because when all nodes in $\PlayerR(v)$ have increase their phases, $\psi$ decreases for all $v$'s  neighbors.
\end{proof}

\begin{lemma}
Algorithm $\AlgoC$ requires $O(\max\{\log \Delta, \log \log n\})$ bits of memory per node.
\label{lem:MemColor}
\end{lemma}

\begin{proof}
  The variables $\delta_v,\couleur_{v},\conflit_{v},\Mc_{v}$ take $O(\log \Delta)$ bits. The compact identifier $\PactID_v$ takes $O(\log \log n)$ bits per node.
\end{proof}

\begin{proof}[Proof of Theorem~\ref{theo:color}]
Direct by Lemma~\ref{lem:Delta}, Lemma~\ref{lem:DecColor}  and Lemma~\ref{lem:MemColor}.
\end{proof}

\begin{lemma}
Algorithm  \AlgoC\/ converges in $O(\Delta^{\Delta^2}n^3)$ steps.
\end{lemma}
\begin{proof}
Direct by the potential function $\Phi(\gamma)$.  
\end{proof}

\subsection{Cleaning a cycle or an impostor-rooted spanning tree}
\label{annexe:Clean}
We now present \AlgoFreeze\/ in Algorithm~\ref{Algo:Freeze}. This algorithm uses only one binary variable $\frozen$. 
 This approach presents several advantages. After $v$ detecting a cycle, the cycle is broken ($v$ deletes its parent), and a frozen node cannot reach its own subtree, due to the cleaning process taking place from the leaves to the root. So, two cleaning processes cannot create a livelock.
 
\LinesNumberedHidden 
\begin{algorithm*}
 \begin{small}
\caption{ Algorithm $\AlgoFreeze$ \label{Algo:Freeze}}
\[
\hspace*{-1cm}
\begin{array}{lcllll}
\RErCycle&:&\ErCycle(v)\vee \ErST(v) &\longrightarrow \frozen_v:=1, \p_v:=\emptyset;\\
\RFroze &:& \neg\ErCycle(v)\wedge \neg \ErST(v)\wedge (\frozen_{\p_v}=1) \wedge (\frozen_{v}=0)&\longrightarrow \frozen_v:=1;\\
\RPrun &:& \neg\ErCycle(v)\wedge \neg \ErST(v)\wedge (\frozen_{\p_v}=1) \wedge (\frozen_{v}=1) \wedge (\Child(v)=\emptyset)&\longrightarrow Reset(v);\\
\end{array}
\]
\end{small}
\end{algorithm*}

\begin{theorem}
\label{theo:Freeze}
Algorithm \AlgoFreeze\/ deletes  a cycle or an impostor-rooted sub spanning tree in $n$-nodes graph in a silent self-stabi\-li\-zing manner, assuming the state model, and a distributed unfair scheduler. Moreover, Algorithm \AlgoFreeze\/  uses $O(1)$ bits of memory per node. 
\end{theorem}
\begin{lemma}
Algorithm \AlgoFreeze\/ converges in $O(n)$ steps. 
\label{lem:freezStep}
\end{lemma}

Proofs of Theorem~\ref{theo:Freeze} and Lemma~\ref{lem:freezStep} see article~\cite{BlinT17}.

\subsection{Self-stabilizing Cycle Detection}

\subsubsection{Correctness of the algorithm $\AlgoBreakR$}

Let  $I_{i}$ be the nearest descendant of $v$ ($I_{i}\neq v$) such that $\minId_{I_{i}}=i$, if such a node exists. And let us denote by $D(v,i)$ the set of nodes on the path between $v$ and $I_{i}$. We suppose that every node $u$ in $D(v,i)$ has $\ErCycle(u)=false$. The value $i$ can improve the value $\minId_{v}$ if and only every node $u$ in $D(v,i)$ has a  $\minId_{u}>i$ and $\del_{u}\neq i$. Also, if $\del_{u}= i$, the value vanishes during the execution, note that may be $u=v$. Predicate $\Improve$ captures this fact.

\begin{equation}
\begin{split}
 \Improve(v,i)\equiv \big(\forall u \in D(v,i): (\minId_{u}>i)\big) \wedge \big[\big(\forall u \in D(v,i):\del_{u}\neq i) \big) \vee \\ \big(\exists u \in D(v,i): (\del_{u}= i) \wedge \Improve(u,j)\big)\big]
\end{split}
\label{pre:Improve}
\end{equation}

Let $\alpha: \Gamma\times V \rightarrow \mathbb{N}$ be the function defined by:
$$\alpha(\gamma,v,i)= 
\left\{
  \begin{array}{ll}
   4 &\text{if } (\minId_{v}>i)\wedge (\del_{v}\neq i) \wedge \Improve(v,i)\\
   3 &\text{if } (\minId_{v}=i) \wedge (\del_{v}\neq i)\wedge  \Improve(v,i) \\
   2 &\text{if } (\minId_{v}=i) \wedge (\del_{v}= i) \wedge  \Improve(v,i)\\
   1 &\text{if } (\minId_{v}> i) \wedge (\del_{v}= i) \wedge  \Improve(v,i)\\
   0&\text{if } \neg \Improve(v,i) \\
  \end{array}
\right.
$$

Let $\beta: \Gamma\times V \rightarrow\mathbb{N}$ be the function defined by:
$$\beta(\gamma,i)=\sum_{v\in V}\alpha(\gamma,v,i)$$

Let $\Xi: \Gamma \rightarrow\mathbb{N}$ be the function defined by:
$$\Xi(\gamma)=(\beta(\gamma,0),\beta(\gamma,1),\dots,\beta(\gamma,IdMax))$$
The comparaison between two configurations $\Xi(\gamma)$ and $\Xi(\gamma')$ is performed using lexical order. 
In the following, $\minId_{v}(\gamma)$ denotes the variable $\minId_{v}$ in configuration $\gamma$.
Note that the algorithm is stabilized when no value $i$ can improve the value stored in $\minId_v$, that is  when $\Xi(\gamma)=0$. We define 
$$\Gamma_{\cal B}=\{\gamma\in \Gamma:\Xi(\gamma)=0\}$$

\begin{lemma}
$true \triangleright \Gamma_{\cal B}$ and $\Gamma_{\cal B}$ is closed. 
\end{lemma}
\begin{proof}
\begin{itemize}
\setlength\itemsep{-0.3em}
\item Rule $\RStart(v)$: $\minId_{v}(\gamma)=\minId_{v}(\gamma')$, so for $i<\minId_{v}(\gamma')$, we have $\beta(\gamma',i)=\beta(\gamma,i)$. Note that for $i>\minId_{v}(\gamma)$, $v$  has no effect on $\beta(\gamma,i)$ and $\beta(\gamma',i)$. Now,  $\beta(\gamma,\minId_{v})=3$ because $\RStart(v)$ is enabled for $v$ only if $(\del_{v}\neq \minId_{v})$, and $\beta(\gamma',\minId_{v})=2$ because we have $(\del_{v}= \minId_{v})$, thus $\beta(\gamma,\minId_{v})=3>\beta(\gamma',\minId_{v})=2$. So, if the scheduler activates $v$ with rule $\RStart(v)$, we obtain $\Xi(\gamma')<\Xi(\gamma)$.
\item Rule $\RID$: 
\begin{itemize}
\item \textbf{$i<\minId_{v}(\gamma)$}: $\beta(\gamma',i)=\beta(\gamma,i)$ because $\ID_v>\minId_{v}(\gamma)$ (otherwise an error is detected). Also, if $i$ can improve $\minId_{v}(\gamma)$, it can also improve $\minId_{v}(\gamma')$.
\item \textbf{$\minId_{v}(\gamma)$:}  Rule $\RID$ needs $\del_{v}(\gamma)=\minId_{v}(\gamma)$, so $\beta(\gamma,\minId_{v}(\gamma))=2$. Now,  we have $\del_{v}(\gamma')= \minId_{v}(\gamma)\neq \minId_{v}(\gamma')$, and we obtain $\beta(\gamma,\minId_{v}(\gamma))=2>\beta(\gamma',\minId_{v}(\gamma))=1$.
\end{itemize}
As a consequence, $\beta(\gamma',i)=\beta(\gamma,i)$ for $i<\minId_{v}(\gamma)$ and $\beta(\gamma',\minId_{v}(\gamma))<\beta(\gamma,\minId_{v}(\gamma))$. So, if the scheduler activates only $v$ for rule $\RID(v)$, we obtain $\Xi(\gamma')<\Xi(\gamma)$.

\item Rule $\RMin(v)$: $\minId_{v}(\gamma')<\minId_{v}(\gamma)$ and $\del_{v}(\gamma)\neq  \minId_{v}(\gamma')$ by definition of rule $\RMin(v)$, so in $\gamma$, we have $\alpha(\gamma,v,\minId_{v}(\gamma'))=4$ because   $\minId_{v}(\gamma')<\minId_{v}(\gamma)$ and $\minId_{v}(\gamma')$ can improve $\minId_{v}(\gamma)$.  Now, we have $\alpha(\gamma',v,\minId_{v}(\gamma'))=3$ because $\del_{v}(\gamma')\neq  \minId_{v}(\gamma')$. $\minId_{v}(\gamma')<\minId_{v}(\gamma)$, so if the scheduler activates only $v$ for rule $\RMin(v)$, we obtain $\Xi(\gamma')<\Xi(\gamma)$.
\end{itemize}
To conclude, $\Xi(\gamma')<\Xi(\gamma)$ for every configurations $\gamma$ and $\gamma'$, when $\gamma'$ occurs later than $\gamma$.
\end{proof}
\begin{proof}[Proof of Theorem~\ref{theo:CycleR}]
Now we prove that, in $\Gamma_{\cal B}$ if the spanning structure $S$ contains a least one cycle $\tt{C}$, then at least one node $v$ in $\tt{C}$ has $\ErCycle(v)=true$. 
For the purpose of contradiction, let us assume the opposite. Let $\gamma\in\Gamma_{\cal B}$ and every node $v$ in the cycle $\tt{C}$ in $\gamma$ has $\ErCycle(v)=false$. By definition  all the nodes in $\tt{C}$ have a parent, and all the nodes have $\minId_{v}\geq \minId_{\kid_{v}}$. Now, if a node $v$ shares the same $\minId$ with its parent and its child, then rule $\RStart$ is enabled for $v$, a contradiction with $\Xi(\gamma)=0$. In a cycle, it is not possible that all nodes have $\minId_{v}> \minId_{\kid_{v}}$ (due to well foundedness of integers, at least one node $v$ has $\minId_{v}< \minId_{\kid_{v}}$), which is a contradiction with the assumption that every node $v$ is such that $\ErCycle(v)=false$.
\end{proof}

\begin{lemma}
Algorithm \AlgoBreakR\/ converges in $O(n^n)$ steps.
\end{lemma}
\begin{proof}
Direct by the potential function $\Xi(\gamma)$.  
\end{proof}

\subsubsection{Algorithm \CAlgoBreakG\/ et Predicates}
Predicate $\ErCycle$ must be updated to take into account this extra care. We denote by $\kid_{v}$ the child of $v$ with minimum compact identifier stored in $\minId_{v}$. Moreover, predicate $\ErCycle$ now takes into account the error(s) related to compact identifiers management (see Equation~\ref{eq:ErT}: $\ErT(v)$). It is important to note that the cycle breaking algorithm does not manage phase differences.  Indeed, a node $v$ with a phase bigger than the phase of one of its children takes the $\minId$ of its children, if its phase its bigger than two or if no child has its same compact identifier.
The compact identifier stored in  $\minId_{v}$ is be compared using lexicographic order by rule $\RMin$.
\begin{equation}
\begin{split}
\ErCycle(v)\equiv (\p_v\neq \emptyset) \wedge \Big((\minId_{\kid(v)}=_c\PactID_v^{\,i})\vee   (\exists (u,w)\in \Child(v): \minId_{u}=_{c}\minId_{w}) \vee (\minId_{v}>_{c}\PactID_v^{\,i})\vee\\ \big( (\minId_{v}\not \simeq_{c} \PactID_v)\wedge (\minId_{v}<_{c}\minId_{\kid(v)})\big) \vee (\Active_v \wedge \ErT(v))\Big)
\end{split}
\end{equation}
\LinesNumberedHidden 
\begin{algorithm*}

\caption{ Algorithm \CAlgoBreakG\/ For node $v$ with  $\neg\ErCycle(v)$}
\[
\hspace*{-1cm}
\begin{array}{lcllll}
\RInc&\sep&\Active_v \wedge (\minId_{\p_{v}}\simeq_{c}\minId_{v}\simeq_{c}\minId_{\kid(v)})&\longrightarrow\SIncPh(v);\\
\RStart&\sep&  \neg \Active_v \wedge (\minId_{\p_{v}}\simeq_{c}\minId_{v}\simeq_{c}\minId_{\kid(v)})\wedge (\del_v\not \simeq_{c} \minId_{v})&\longrightarrow \del_v:=\minId_{v};\\
\RMin&\sep& (\minId_{v}>_c\minId_{\kid(v)}) \wedge (\del_{v}\not \simeq_{c} \minId_{\kid(v)})&\longrightarrow \minId_v:=\minId_{\kid(v)}, \Active_v:=false;\\
\RID&\sep& (\del_{\p_{v}}\simeq_{c}\del_{v}\simeq_{c}\del_{\kid(v)}\simeq_{c}\minId_{v})\wedge (\minId_{v}\not \simeq_{c} \PactID_{v}^{\,1})&\longrightarrow \minId_{v}:=\PactID_v^{\,1},\Active_v:=true;\\
\end{array}
\]

\end{algorithm*}

The proof of theorem~\ref{theo:CCycle}  mimics the proof of algorithm $\AlgoBreakR$.
\begin{lemma}
Algorithm \CAlgoBreakG\/ converges in $O(n^n \log n)$ steps.
\end{lemma}
\subsection{Spanning Tree Construction without Distance to the Root Maintenance}
\subsubsection{Predicates}
The function $\Cand(v)$ returns the color of the neighbor of $v$ with the maximum root:
 \begin{equation}
\Cand(v)=\{\couleur_{u}:u\in N(v) \wedge \R_{u}=\max\{\R_{w}:w\in N(v)\}\} 
\end{equation}

We now present a list of trivial errors and impostor-root errors for the construction of the spanning tree. The explanations of the different elements composing the predicate $\ErST(v)$ follow:
(1) A node without relative has its root signature different to its own variables.
(2) The variable $\degr_v$ is not equal at the degree of $v$.
(3) The invariant is not satisfied.
(4) A node with $\new_v= \emptyset$ (that is, $v$ is not involved in a rerouting process) has a root signature bigger than that of its parent;
(5) A node with $\new_v\neq \emptyset$ (that is, $v$ is  involved in a rerouting process) has a root signature different from that of its tentative new parent.
(6) A root $v$ with $\new_v= \emptyset$ and a signature $\R_{v}$ that does not match is own.
(7)A node involved in a rerouting process whose parent's parent is itself.

\begin{equation}
\ErST(v)\equiv
\left\{
\begin{array}{ll}
(1)&\big((\p_v=\emptyset)\wedge (\Child(v)=\emptyset) \wedge [(\R_v\neq(\deg_v,\couleur_v,\ID_v)) \vee  (\new_v\neq\emptyset)]\big)\vee\\
(2)&(\degr_v\neq \deg_v)  \vee \\
(3)&(\N(v)\not \in \{\{\p_v\}\cup\Child(v)\}\}) \vee \\
(4)&\big((\new_v= \emptyset) \wedge (\R_v> \R_{\p_v}) \big) \vee \ \\
(5)&\big((\new_v\neq \emptyset) \wedge (\R_v\neq \R_{\new_{v}})\big) \vee \\
(6)& \big((\p_v=\emptyset) \wedge (\new_v= \emptyset) \wedge (\R_v\neq(\deg_v,\couleur_v,\ID_v))\big)\vee\\
(7)&\big((\new_v\neq \emptyset) \wedge (\p_{\p_v}=\couleur_v)\\
\end{array}
\right.
\end{equation}

\subsubsection{Spanning Tree Construction without Distance to the Root Maintenance}

Proof of Theorem~\ref{theo:ST}
Let $\N(v)$ denote the color of $v$'s neighbor with the maximum degree, and in case there are several such neighbors, the one with maximum color.
\begin{equation}
\N(v)=\{\couleur_u: u\in N(v)\wedge \degr_u=\max\{\degr_w:w\in N(v)\} \wedge \couleur_u=\max\{\couleur_w:w\in N \wedge (\degr_w=\degr_u)\}\}
\end{equation}

\begin{lemma}[Invariant]
For every node $v\in V$ such that $\p_{v}\neq\emptyset$ and $ \Child(v)\neq \emptyset$, $\N(v)\in \{\p_{v}\cup \Child(v)\}\neq \emptyset$ remains true. 
\label{lem:invariant}
\end{lemma}

\begin{proof}
Proof by induction
\begin{description}
\item[Basis case:]
When a node $v$ starts the algorithm it chooses for a parent the node with the maximum degree, if there exist more than one it chooses the one with the maximum identifier  among the ones with the maximum degree. So if a node picks a parent $u$ at the first execution of the algorithm it takes $u= \N(v)$ so for theses nodes  the invariant is preserved. 
For the nodes $v$ which are a maximum local. We denote by $u$ the node $\N(v)$. Suppose that at the first execution of $u$, $u$ choses  the node $w$ as a parent, that means $\R_w>\R_u$ and $\R_{w}>\R_{v}$. So after this execution $\R_u=\R_w$, so now $v$ can choose $u$ as a parent and the invariant is preserved for node $v$. So after one execution of the algorithm for all the nodes the invariant is preserved. 
\item[Assumption:] Assume true that after $t$ steps of execution, the algorithm   preserved the invariant.
\item[Inductive step:] Let us consider a node $v$, by the assumption we have 
$$\N(v)\cap\{ \{\p_{v}\}\cup \Child(v)\}\neq \emptyset$$
The node $v$ cannot change its children in can only change its parent, and only if $v$ its a root (see rules $\RMerge$ and $\RReRoot$ of  Algorithm~\ref{Algo:ST}). So for $v$ $\N(v)\in \Child(v)$, the rule $\RMerge$ assigns as a parent a new neighbor $u$ ($u\not \in \Child(v)$) so the invariant is preserved.
The rule $\RReRoot$ assigns as a parent of $v$ a child of $v$ so  the invariant is preserved.
\end{description}
\end{proof}

\begin{lemma}
The descendants $u$ of $v$ with $\new_u=\emptyset$ have $\R_u\leq \R_v$.
\label{lemma:descendants}
\end{lemma}
\begin{proof}
Proof by induction on the value $\R_u$ with $u$ descendants of $v$

\begin{description}
\item[Base case:] Each node $v\in V$ with $\p_v=\emptyset$ and $\Child(v)=\emptyset$ has $\R_v=(\deg_v,\couleur_v,\ID_v)$ and $\new_v=\emptyset$, otherwise an error is detected.  
A node $v$ takes a parent iff there exists a neighbor $w$ of $v$ such that $\R_w>\R_v$, and in this case $v$ maintains its variable $\new_w=\emptyset$, so the claim is satisfied (see Rule $\RMerge$).
\item[Assumption:] Assume that there exists a configuration $\gamma$ where for every node $v\in V$, all the descendants $u$ of $v$ with $\new_u=\emptyset$ have $\R_u\leq \R_v$.
\item[Inductive step:] We consider Configuration $\gamma+1$. For a node $v$ and every descendants $u$, the assumption gives the property that if $\new_u=\emptyset$, then $\R_u\leq \R_v$. 
Let us now consider the case where there exists a neighbor  $w$  of $u$ with  $\R_u<\R_w$.  
\begin{smallitemize}
\item If $w$ is the parent of $u$, $\R_u$ takes the value of $\R_w$ (see rule $\RUpdate$). By the induction assumption, we have  $\R_w\leq\R_v$ (as a parent of $u$, $w$ is also a descendant of $v$). So, $\R_u$ remains inferior or equal to $\R_v$.
\item By the induction assumption, if $\R_w>\R_u$, then $w$ cannot be a descendant of $u$.
\item If  $w$ is not in the same subtree of $u$, $u$ cannot change its  parent because $u$ is not a root (see rule $\RMerge$). So $u$ changes its $\R_u$ to $\R_w$, but it sets  $\new_u=w$ (see rule $\RPath$).
\end{smallitemize}
Now, if there exists a neighbor $w$ of $u$ such that $\R_w=\R_u$, then to execute rule $\RReRoot$, $u$ must be a root. We obtain a contradiction with our assumption that $u$ is a descendant of $v$.

To conclude, if $u$ is the descendant of $v$ in configuration $\gamma$ and it remains a descendant of $v$ at configuration  $\gamma+1$, and the value of $\new_u$ remains empty,  then $\R_u\leq \R_v$ in configuration  $\gamma+1$.
\end{description} 
\end{proof}

\begin{lemma}
If there exists an acyclic spanning structure in Configuration $\gamma$, then any execution of a rule maintains an acyclic spanning structure in Configuration $\gamma+1$. \footnote{When a node $v$ has $\p_v\in \Child(v)$, we delete $\p_v$ (see rule~$\RDel$), so we do not consider this case as a cycle.} 
\label{lemma:acyclic}
\end{lemma}

\begin{proof}  Proof by induction on the size of the acyclique spanning structure.
\begin{description}
\item[Basis case:] By contradiction: Remark that, thanks to Algorithm $\AlgoC$, there exist a total order between the neighbors of a node. Let us consider three neighbor nodes $a,b,c\in V$ such that in Configuration $\gamma$, $a,b$ and $c$ have no relatives. Then all three nodes are enabled by rule~$\RMerge$. Let us suppose for the purpose of contradiction that in Configuration $\gamma+1$ a cycle exists. More precisely: $\p_{a}=b$, $\p_{b}=c$ and $\p_{c}=a$, to achieve that :
\begin{enumerate}
\setlength\itemsep{-0.1em}
\item  $a$ must choose $b$ as a parent, for that $\R_{b}>\R_{c}$ 
\item $b$ must choose $c$ as a parent, for that $\R_{c}>\R_{a}$ 
\item $c$ must choose $a$ as a parent, for that $\R_{a}>\R_{b}$ 
\end{enumerate}
We obtain a contradiction between (1),(2) and (3).

\item[Assumption:] Assume true that in configuration $\gamma$ there exists an acyclic spanning structure.  
\item[Inductive step:] 
  Let us consider a node $v\in V$,  a node $v$ takes a new parent only in two cases, and in both case $v$ must be a root.

Let us consider first rule~$\RMerge$, let $u$ be the neighbor of $v$ with $\R_v<\R_u$ and $\new_u=\emptyset$. By Lemma~\ref{lemma:descendants}, $u$ is not a descendant of $v$, so if $v$ takes $u$ as a new parent, an acyclic spanning structure is preserved. 
Now, we consider rule~$\RReRoot$. Let $u$ be the neighbor of $v$ such that $\R_u=\R_v$ and $\new_v=u$. In this case, $u$ is either a child of $v$ with $\new_u\neq \emptyset$, or $v$ is not a child of $v$ with $\new_u=\emptyset$. If $v$ is a child of $v$, $v$ takes $u$ as a parent. Remark that the first action of $u$ is  to delete its parent (see rule $\RDel$, and consider the fact that all other rules require $\p_{v}\not \in \Child(v)$), so we do not consider this case as a cycle.  If $v$  is not a child of $v$ with $\new_u=\emptyset$, by Lemma~\ref{lemma:descendants} $u$ is not a descendant of $v$. Now, when $v$ takes $u$ as a parent, this action maintains an acyclic spanning structure.

To conclude, Configuration $\gamma+1$ maintains a acyclic spanning structure.
\end{description}
\end{proof}

\begin{lemma}
If $v$ is a node such that $\R_{v}=\R_{r}$, and every ancestor of $v$ (and $v$ itself) have $\new=\emptyset$. Then $r$ is an ancestor of $v$, or $v$ itself. 
\label{lemma:ancestor}
\end{lemma}

\begin{proof}
Suppose for the purpose of contradiction that $r\neq v$, and $r$ is not an ancestor of $v$. By Lemma~\ref{lemma:acyclic}, $v$ is an element of a sub spanning tree $\T$. Let $w$ be the oldest ancestor of $v$ such that $\new_{w}=\emptyset$. By hypothesis, every ancestor $z$ of $v$ (including $w$) has $\R_{z}\neq \R_{r}$. By Lemma~\ref{lemma:descendants}, we have $\R_{v}\leq \R_{z}$, which contradicts $\R_{v}=\R_{r}$.
\end{proof}

\begin{lemma}
Executing Algorithm $\AlgoST$ constructs a spanning tree rooted in node with the maximum degree, maximum color, maximum identifier, assuming the state model, and a distributed unfair scheduler.
\end{lemma}

\begin{proof}
Let $\psi: \Gamma\times V \rightarrow \mathbb{N}$ be the function defined by:
$$\psi(\gamma,v)=  \big((\deg_{\ell}-\degr_{v})+(\couleur_{\ell}-\couleur_{v})+(\ID_{\ell}-\ID_v)\big) $$
where $\ell$ is the node with $\R_{\ell}>\R_v$ with $v\in V\setminus\{\ell\}$.
Now, let $\phi: \Gamma\times V \rightarrow \mathbb{N}$ be the function defined by:
$$\phi(\gamma,v)=
\left\{
  \begin{array}{ll}
   2 &\text{if } \new_v\neq \emptyset\\
  1 & \text{if } \p_{\p_v}=\couleur_v\\
     0 &\text{otherwise} \\
  \end{array}
\right.
$$
Remark that a node $v$ cannot have $\new_v\neq \emptyset$ and $\p_v\in \Child(v)$. Otherwise an error is detected through predicate $\ErST(v)$.

Let $\Psi: \Gamma \rightarrow\mathbb{N}$ be the potential function defined by:
$$\Psi(\gamma)=\Big(\sum_{v\in V}\psi(\gamma,v),\sum_{v\in V} \phi(\gamma,v)\Big).$$ 

Let $\gamma$ be a configuration such that $\Psi(\gamma)>0$, and let $v$ be a node in $V$ such that $v$ is enabled by a rule of Algorithm  $\AlgoST$. 
If $v$ executes rules $\RUpdate$, $\RMerge$, or $\RPath$,  then  $\R_{v}$ increases and  we obtain $\psi(\gamma',v)<\psi(\gamma,v)$. Now, if $v$ executes rule 
$\RReRoot$, this implies $\new_v$ is not empty. After execution of $\RReRoot$, $\new_v$ become empty, so $\phi(\gamma,v)$ decreases by one. Finally, if $v$ executes rule $\RDel$, it implies that $v$ had $\p_{\p_v}=\couleur_v$, and now $\p_v=\emptyset$. As a result, $\phi(\gamma',v)=\phi(\gamma,v)-1=0$.

Therefore, we obtain $\Psi(\gamma')<\Psi(\gamma)$. By Lemmas~\ref{lemma:acyclic} and \ref{lemma:ancestor} we obtain the property that when $\Psi(\gamma)=0$, a spanning tree rooted in $\ell$ is constructed.
\end{proof}

\begin{lemma}
Algorithm  \AlgoST\/ converges in $O(\Delta n^3)$ steps.
\end{lemma}
\begin{proof}
Direct by the potential function $\Psi(\gamma)$.  
\end{proof}
\begin{lemma}
Algorithm  \AlgoCST\/ converges in $O(\Delta n^3\log n)$ steps.
\end{lemma}

The proof of Theorem~\ref{theo:CST} mimics the proof of Theorem~\ref{theo:ST}.

\subsection{Corrects of the algorithm of Leader Election}
\begin{proof}[Proof of Theorem \ref{theo:CompoC}] 
  We first need to show that the number of activations of rules of algorithm $\AlgoCST$ are bounded if there exist nodes enabled by $\AlgoC$, $\AlgoFreeze$ or $\CAlgoBreakG$.  Let us consider a subset of the nodes $A$ enabled for at least one of these algorithms, and  by $S$ the nodes enabled by rule $\AlgoCST$. The nodes in $S$ belong to some spanning trees (possibly only one), otherwise at least one of rules of $\AlgoFreeze$ or $\CAlgoBreakG$ would be enabled. So, there exist a node in $S$ that is enabled by algorithm $\AlgoCST$. Algorithm $\AlgoCST$ is talkative, but it runs by waves, and its waves require that all neighbors of a node $v$ have the same $\R$ at each phase. As we consider connected graphs only, there exists at least one node $v$ in $S$ with a neighbor $u$ in $A$. Then, there exists a configuration $\gamma'$ where the rules of $\AlgoCST$ are not enabled, because $u$ cannot have the same $\R$ at each phase (since $u$ is not enabled by rules of $\AlgoCST$).
So, only Algorithms $\AlgoC$, $\AlgoFreeze$, and $\CAlgoBreakG$ may now be scheduled for execution, as they have higher priority. As they are silent and operate under an unfair distributed deamon, we obtain convergence.

Let us now consider a configuration $\gamma$ where no node are enabled for Algorithm $\AlgoC$, $\AlgoFreeze$, and $\CAlgoBreakG$. Yet, there exists a node enabled by Algorithm  $\AlgoCST$. Thanks to Theorem~\ref{theo:CST}, we obtain a spanning tree rooted in the node with the maximum degree, maximum color, and maximum identifier. As a consequence, only the root $r$ has $\ell_r=true$ and every other node $v\in V\setminus\{r\}$ has $\ell_v=false$.
\end{proof}

\end{document}